# Advancing Alzheimer's disease treatment: Synergistic ligand combinations targeting BACE1 through multi-ligand simultaneous docking

**Pronama Biswas[1†*], Surya Shanbhog[2], Merla Sudha[1], and Belaguppa Manjunath Ashwin Desai[3†]**

[1]Department of Biological Sciences, School of Basic and Applied Sciences, Dayananda Sagar University, Kudlu Gate, Hosur Road, Bengaluru, India

[2]AIIMS–Bathinda, Mandi Dabwali Rd, Bathinda, Punjab, India

[3]Department of Electronics and Communication Engineering, School of Engineering, Dayananda Sagar University, Devarakaggalahalli, Harohalli, Kanakapura Road, Ramanagara Dt., Bengaluru, India

†*These authors contributed equally to this work.*

***Corresponding author:** Pronama Biswas (pronamabiswas@gmail.com; pronama-sbas@dsu.edu.in)

## Abstract

Alzheimer's disease, a progressive neurodegenerative disorder, is characterized by memory loss, cognitive decline, and behavioral changes, with no therapies available to halt its progression. Newer therapies targeting BACE1, a key enzyme in amyloid plaque formation, have shown promise in clinical trials. However, they have been limited by side effects and insufficient efficacy in slowing cognitive decline. Current treatments, including phase III BACE1 inhibitors such as atabecestat, elenbecestat, lanabecestat, and verubecestat, have shown limited efficacy. This is the first study where multi-ligand simultaneous docking (MLSD) was employed to identify potential synergistic inhibitor combinations from among thousands of small molecules that could yield better results than the current phase III inhibitors of BACE1. A library of 15,641 small molecules with known $IC_{50}$ values against BACE1 was filtered and docked, yielding binding affinities from −11.32 kcal/mol to +10.85 kcal/mol. Five small molecules with affinities ≤−11 kcal/mol and phase III inhibitors were tested as combinations for MLSD. Among 14 combinations tested, CHEMBL4078427 and CHEMBL3656158, CHEMBL4078427 and CHEMBL3695732, verubecestat and CHEMBL3656158, and CHEMBL4078427 and lanabecestat demonstrated superior binding affinities of −19.90 kcal/mol, −18.45 kcal/mol, −18.07 kcal/mol, and −17.67 kcal/mol, respectively, with inter-ligand interactions indicating synergy. Molecular dynamics simulations of CHEMBL4078427 with lanabecestat revealed enhanced BACE1 inhibition, showing lower root mean square deviation and radius of gyration compared to single-ligand inhibition. These findings underscore the transformative potential of MLSD in identifying synergistic compound interactions, paving the way for novel combination therapies in Alzheimer's disease treatment.

## 1. Introduction

Alzheimer's disease (AD) is a progressive neurodegenerative disorder marked by dementia and cognitive decline affecting speech, personality, judgment, vision, and daily functioning.[1] It is the leading cause of dementia, accounting for ~70% of cases, and affected an estimated 55 million people globally in 2019, a number projected to rise to 139 million by 2050, mainly in low- and middle-income countries.[2] AD is a major cause of disability and mortality in the elderly, second only to vascular and malignant diseases,[3] with global costs expected to reach $2.8 trillion by 2030.[4]

The complex pathophysiology of AD challenges understanding and treatment development. Key features include extracellular amyloid-β (Aβ) plaques and intracellular tau tangles linked to disease progression. AD pathogenesis involves enzymatic cleavage of the amyloid precursor protein (APP) by β-secretase (BACE1) and γ-secretase produces Aβ40 and Aβ42 peptides, which are hydrophobic, prone to aggregation, and form neurotoxic plaques.[5] These peptides also trigger tau hyperphosphorylation, disrupting microtubules and neuronal function, leading to memory loss (Figure 1).[6] γ-Secretase is a membrane-bound protease composed of presenilin, nicastrin, APH-1, and PEN-2,[7] while BACE1 is primarily neuronal and active in endosomes, vesicles, and Golgi compartments. Inhibiting these enzymes shows therapeutic potential in AD.

γ-Secretase plays a crucial role in NOTCH signaling, making γ-secretase inhibitors (GSIs) prone to severe adverse effects.[8-10] BACE1 inhibition has thus emerged as a disease-modifying strategy with the potential to slow AD progression (Figure 1). In animal models, BACE1 inhibitors reversed amyloid pathology.[11] BACE1 is highly expressed in the brain, regulating neuronal functions, while its activity in other tissues is lower.[12] Notably, ~30% of sporadic AD patients show a two- to fivefold increase in BACE1 levels and a 68.9% rise in plasma activity,[13-15] highlighting the importance of developing BACE1 inhibitors in AD research.[16]

Therapeutic efforts in AD have mainly targeted pathology, but approved treatments such as cholinesterase inhibitors and NMDA receptor antagonists provide only symptomatic relief and lack disease-modifying effects.[17] These drugs also cause severe side effects, underscoring the need for novel strategies.[18] Small molecules are promising due to their favorable pharmacokinetics, including brain penetration and high permeability. As non-peptide organic compounds under 1000 Daltons, they can be readily modified to enhance pharmacokinetic and pharmacodynamic properties, improving binding affinity and specificity for AD-related targets.[19,20]

The failure rate of AD drug candidates in clinical trials is nearly 98%, underscoring the challenges in developing effective therapies.[21] Computational approaches, particularly computer-aided drug design (CADD), enable faster and more efficient screening of drug candidates compared to traditional methods. Techniques such as molecular docking are widely used to estimate ligand binding affinity and have identified inhibitors of multiple AD-related proteins.[22-24] For instance, an *in silico* study by Jabir *et al.*[24] reported PDB Ligand ID: 6Z5 as a potential inhibitor of both β-secretase and γ-secretase.

However, traditional docking that evaluates a single ligand at a time may fail to capture the complex interplay of multiple ligands competing or cooperating at the binding site, leading to limited predictive accuracy. Multi-ligand simultaneous docking (MLSD) addresses this limitation by assessing several ligands with a protein target simultaneously, providing insight into competitive, cooperative, and synergistic interactions. Given the complex pathophysiology, pharmacokinetic demands, and high trial failure rates in AD drug development, MLSD-driven combination strategies offer a promising avenue for discovering effective BACE1 inhibitors.

## 2. Methods

### 2.1. Protein acquisition and validation

The three-dimensional structure of BACE1 was obtained from the Research Collaboratory for Structural Bioinformatics (RCSB) Protein Data Bank (PDB ID: 7B1E). The structure was modeled using X-ray diffraction and exhibited a resolution of 1.62 Å. The quality of the protein structure was evaluated using online tools PROCHECK (https://saves.mbi.ucla.edu) and VoroMQA (https://bioinformatics.lt/wtsam/voromqa). Specifically, the assessment included analyses with Whatcheck, Verify 3D, and Procheck within PROCHECK, aiming for a Ramachandran score exceeding 90%, a Verify 3D value above 80%, and ensuring the VoroMQA Global plot position was above the lowest 5% threshold[25] (Figure 2).

### 2.2. Ligand selection and acquisition

Ligands were selected from the ChEMBL database, with a focus on pre-existing inhibitors that have documented $IC_{50}$ data against BACE1 from wet lab studies (https://www.ebi.ac.uk/chembl/). The selection process involved querying the database for BACE1 under the "All Targets" category, followed by filtering the results to include only those associated with *Homo sapiens* and classified as "SINGLE PROTEIN." Among the available BACE1 entries, CHEMBL4822 was chosen due to its extensive dataset, comprising 15,641 small molecules. To refine the dataset for ligands with a higher likelihood of crossing the blood–brain barrier (BBB), a molecular weight threshold of <500 Daltons was applied. Figure 3 shows the distribution of $pIC_{50}$ values of small molecules. An $IC_{50}$ value of <10 micromole ($pIC_{50}$ >5) was used as a criterion, as lower $IC_{50}$ values are indicative of better pharmacokinetics or increased potency. These filtering steps resulted in the selection of 6,548 compounds for further analysis.

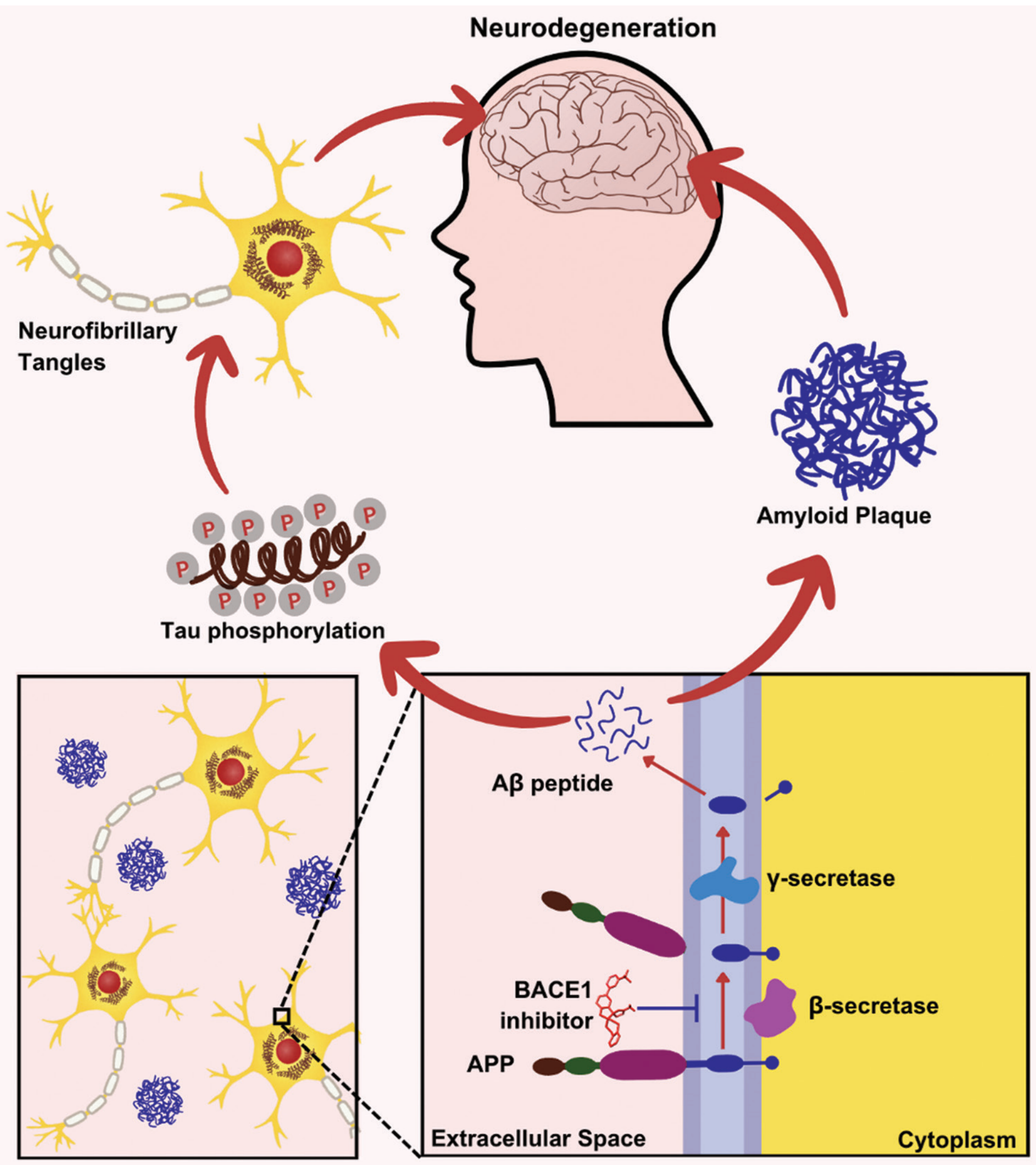

**Figure 1.** Amyloid precursor protein (APP) undergoes sequential cleavage by β-secretase (BACE1) and γ-secretase yielding Aβ peptides that are responsible for the formation of amyloid plaques and microtubule disruption, resulting in the formation of neurofibrillary tangles. This results in the neurodegenerative process associated with Alzheimer's disease. A β-secretase inhibitor is capable of reducing the production of the Aβ peptides. Thus, effectively halting disease progression.

ADMETlab 3.0 (https://admetlab3.scbdd.com/) was used to predict the pharmacokinetic and pharmacodynamic characteristics of the ligands. Ligands with highly unfavorable properties such as solubility, BBB permeability, flagging of TOX21 pathways, and multiple high-level systemic toxicities were filtered. As part of the CADD workflow, PAINS and Brenk filtering were applied to ensure that the screening process yielded reliable, reproducible results by removing compounds likely to cause non-specific or artifact-based effects. Pan-assay interference compounds (PAINS) are a class of compounds known to produce false positives by interfering with the assay readout, not by specifically interacting with the biological target of interest, but through non-specific, often artifact-generating mechanisms.[26] "Brenk-filtering" removes molecules containing substructures associated with undesirable pharmacokinetics or toxicity, such as sulfates and phosphates that contribute to poor pharmacokinetics, nitro groups known for their mutagenic properties, and 2-halo pyridines and thiols, both of which are highly reactive.[27] A final set of 4,501 small molecules was chosen after excluding compounds flagged for PAINS and Brenk using RDKit (https://www.rdkit.org/).

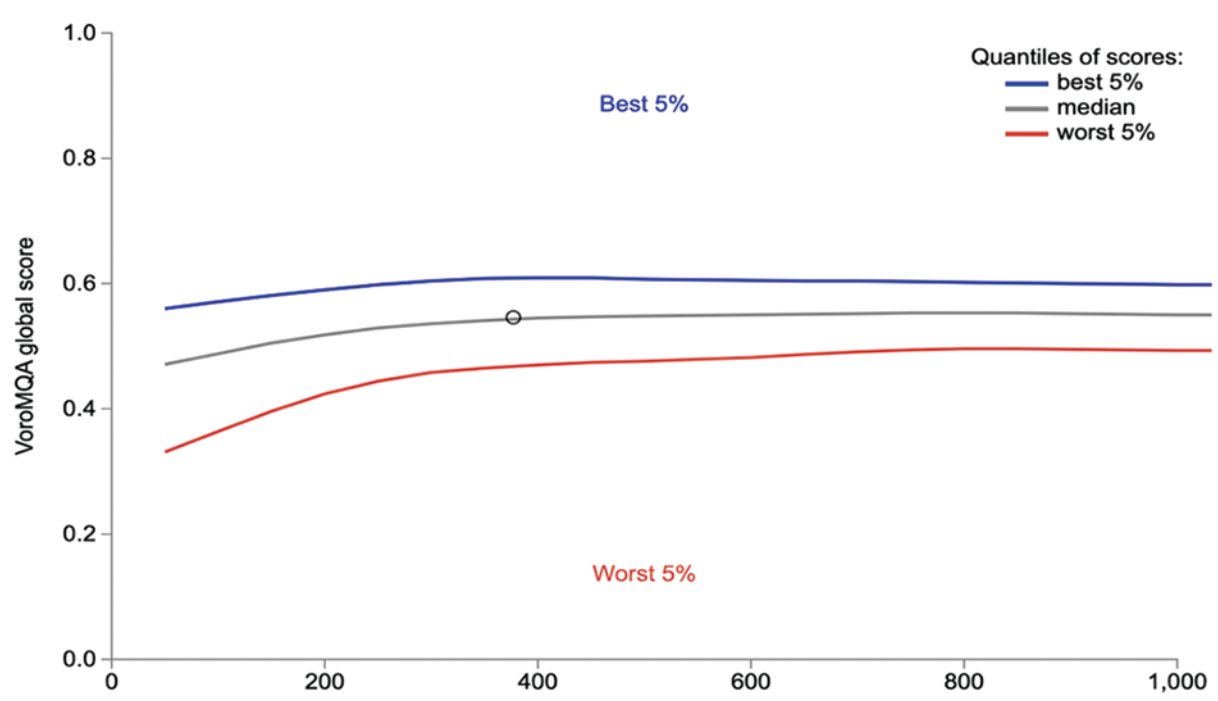

**Figure 2.** VoroMQA Global plot position for the BACE1 protein (PDB ID: 7B1E). The plot position shows that the BACE1 protein model's quality is slightly above the median line (indicated in gray) and well above the lowest 5% threshold (indicated in red). Diagram was obtained from https://bioinformatics.lt/wtsam/voromqa.

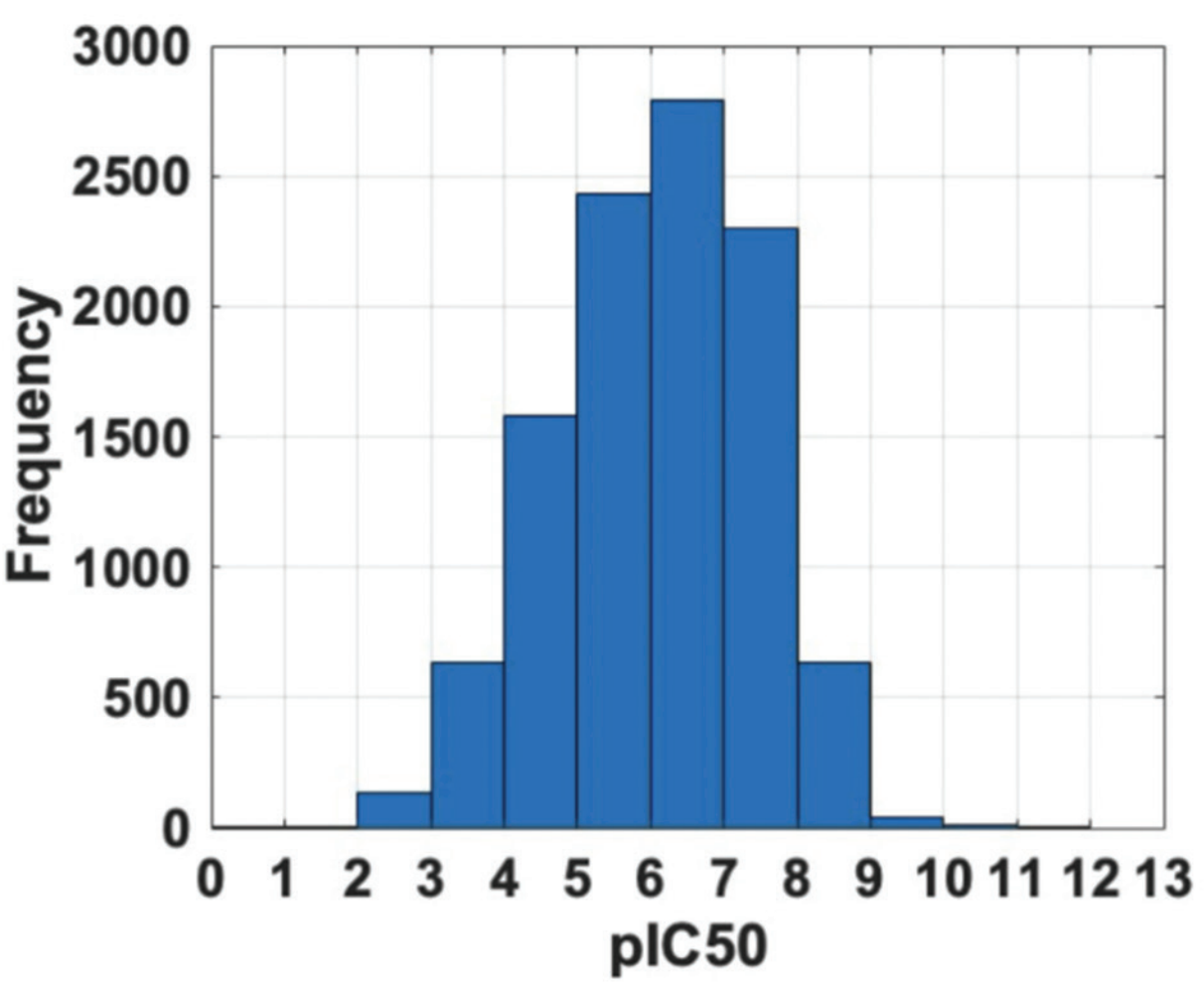

**Figure 3.** Distribution of $IC_{50}$ values of the small molecules against BACE1. $pIC_{50}$ >5 was used as a criterion for filtering the compounds.

The three-dimensional structures of these molecules were obtained by converting Simplified Molecular-Input Line-Entry System (SMILES) notation retrieved from ChEMBL using Open Babel. After PAINS and Brenk filtering, the final set of small molecules was subjected to single-ligand docking, and the top five compounds with binding affinities lower than −11 kcal/mol were selected. Binding affinity values are expressed in terms of Gibbs free energy, where more negative values indicate stronger protein–ligand interactions.[28] Therefore, compounds with affinities below −11 kcal/mol were prioritized for further analysis. For comparison, Saha *et al.*[29] performed a molecular docking study of phytochemicals from *Moringa oleifera* against BCL2 and reported that the strongest binding was observed for niazinin with 10.19 kcal/mol, which was considered a strong interaction. Since −10 kcal/mol has been regarded as a good binding affinity value in prior studies, our cutoff of −11 kcal/mol represents an even more stringent and reliable threshold for selecting potent candidates. This process yielded five final small molecules, which were subsequently analyzed for potential therapeutic combinations using MLSD. In addition, four phase III inhibitors (atabecestat, lanabecestat, elenbecestat, verubecestat) were chosen from the "Drugs and Clinical Candidates" section for CHEMBL4822. Structure Data File (SDF) format files were acquired from PubChem (http://pubchem.ncbi.nlm.nih.gov) by searching for the compounds in the PubChem database, with PubChem Compound IDs 68254185, 67979346, 57827330, and 51352361, respectively. These inhibitors were used for validation and comparative analysis.

## 2.3. Tools and software

Open Babel (http://openbabel.org) was used on the Ubuntu platform to convert ligand SMILES files into Protein Data Bank, Partial Charge, and Atom Type (PDBQT) file format necessary for molecular docking studies. The protein structure was prepared using AutoDockTools 1.5.7, with the processed files saved in PDBQT format. ADMET (Absorption, Distribution, Metabolism, Excretion, and Toxicity) screening was performed using ADMETlab 3.0 (https://admetlab3.scbdd.com/). PAINS and "Brenks" flagging were performed using RDKit (https://www.rdkit.org/). Molecular docking was performed using the Vina Script methodology within a Conda environment accessed through Visual Studio Code. The resultant PDBQT files were subsequently analyzed using. Discovery Studio 2021 Client (https://www.3ds.com/products/biovia/discovery-studio/visualization) and UCSF Chimera 1.17.3 to investigate amino acid interactions and identify binding pockets.[30] Studies were conducted on a server featuring an AMD EPYC 7742 64-Core Processor, 503 GiB of RAM, and NVIDIA A100 GPUs running Ubuntu 22.04.4.

## 2.4. Molecular docking

Computational molecular docking was performed according to the protocol designed by Forli *et al.*[31] BACE1 (PDB ID: 7B1E) was prepared using AutoDockTools 1.5.7. Briefly, the co-crystallized ligand NB-641 (~{N}-[3-[(4~{S})-2-azanyl-4-methyl-5,6-dihydro-1,3-thiazin-4-yl] phenyl]-5-bromanyl-pyridine-2 carboxamide) and water molecules were removed, hydrogen atoms and Gasteiger charges were added, and the missing atoms were repaired. The resulting receptor PDB file was saved in PDBQT format.[32] Ligands were prepared using the "obabel" command in Ubuntu to add hydrogens and Gasteiger charges, followed by energy minimization and conversion to PDBQT files. The exhaustiveness value was set at the default value of 8. Blind docking was performed by configuring the grid box to encompass the entire receptor, with grid dimensions calculated using AutoDockTools. The selected ligands, including phase III drugs, were docked individually with BACE1 using the Vina scoring function within Visual Studio Code. A dedicated Conda environment was established in VS Code to facilitate the docking sessions through batch processing commands.[33]

The resulting output files were sorted based on docking scores, and the top five compounds, with binding affinities below −11 kcal/mol, were selected, along with phase III drugs, for analysis of binding pockets and interactions with amino acids. In MLSD, two ligands were docked simultaneously to the receptor, allowing them to interact with each other and the receptor. During MLSD, the exhaustiveness value was increased to 32 for more consistent docking results. The ligand pairs were selected based on their binding position in the active site of the protein and their interactions with amino acids. For visualization in Biovia Visualizer, ligands were individually selected, copied, and pasted into the 'receptor.pdbqt' tab. The 2D interactions of each ligand were analyzed separately. A similar approach was employed in UCSF Chimera for 3D binding pocket visualization. The schematic representation of the steps followed is illustrated in Figure 4. The protocol for MLSD has been detailed in a study conducted by Sudha *et al.*[25]

Root mean square deviation (RMSD) validation was conducted to validate the procedure and preparation of receptor and ligand molecules and the docking steps. This process involved redocking the co-crystallized ligand of the selected protein using the same methodology described previously, followed by comparing its RMSD with the native binding pose of the co-crystallized ligands in the protein PDB file. UCSF Chimera was employed to superimpose the re-docked ligands and the native poses of co-crystallized ligands and to compare their RMSD values. A lower RMSD value indicates a more accurate docking procedure. Typically, an RMSD value lower than 2 Å is acceptable. This rigorous validation process ensures the reliability and accuracy of the docking procedure, enhancing confidence in the predicted ligand-protein.[25] The co-crystallized ligand of BACE1, NB-641 (identified from RCSB PDB) was used for the RMSD validation. After redocking, RMSD result with the native and redocked pose is 1.016 Å. This can be considered a good RMSD result, thereby verifying our docking protocol.

In addition, the docking results were validated using other docking software. SwissDock running on Attracting Cavities 2.0 was used to dock SM1953 and SM690 with BACE1. The SMILES of the ligands were entered into their web portal. The BACE1 protein target was prepared using AutoDockTools 1.5.7 and uploaded to the web portal. The results showed the SM690 docked within the established binding pocket of BACE1 in 6 out of 10 trials with the highest binding free energy of a ligand approximated by the SwissParam score to be −7.70 kcal/mol. Similarly, another molecular docking pipeline software, MzDock, was also used to cross-validate the findings by blind docking the ligands onto the target protein. Verubecestat docked within the expected binding pocket of BACE1 in 3 out of 3 trials with the highest reported binding affinity of −8.2 kcal/mol. SM1953 also docked successfully within the expected binding pocket of BACE1 in all three trials with the highest reported binding affinity of −10.3 kcal/mol. PyRx is another molecular docking tool that was used to validate our results. Lanabecestat was docked onto BACE1 in this trial. This resulted in lanabecestat successfully docking within the binding pocket in all three trial runs with the highest binding affinity of −9.2 kcal/mol.

### 2.5. Molecular dynamics (MD)

MD simulations were conducted using GROMACS 2024.2 with the CHARMM27 force field for the protein. Ligand topology files were prepared using CGenFF. The docked complexes were solvated in a dodecahedral box with the TIP3P water model, and the system was neutralized by adding sodium and chloride ions. Energy minimization employed the steepest descent algorithm with 50,000 steps. System equilibration was performed under NVT and NPT conditions, each for 100 ps (50,000 steps), maintaining a temperature of 300 K and a pressure of 1 bar. The MD simulation was performed for 100 ns with a 2-fs time step, which was split into five 20 ns segments using checkpoints that were later merged. RMSD and radius of gyration (Rg) were assessed for structural stability and compactness. Free energy calculations were performed on the entire MD trajectory using gmx MMPBSA. In addition, per-residue decomposition analysis was conducted for residues within 4 Å of the ligand, utilizing the leaprc.protein.ff14SB force field to identify key interactions.[25]

## 3. Results

### 3.1. Single ligand docking

Single ligand docking revealed the distribution of binding affinities of the small molecules, as shown in Figure 5, with most values falling in the range of −8 to −9 kcal/mol. Since single ligand docking was the final step in our filtering process, it was crucial to arrive at a

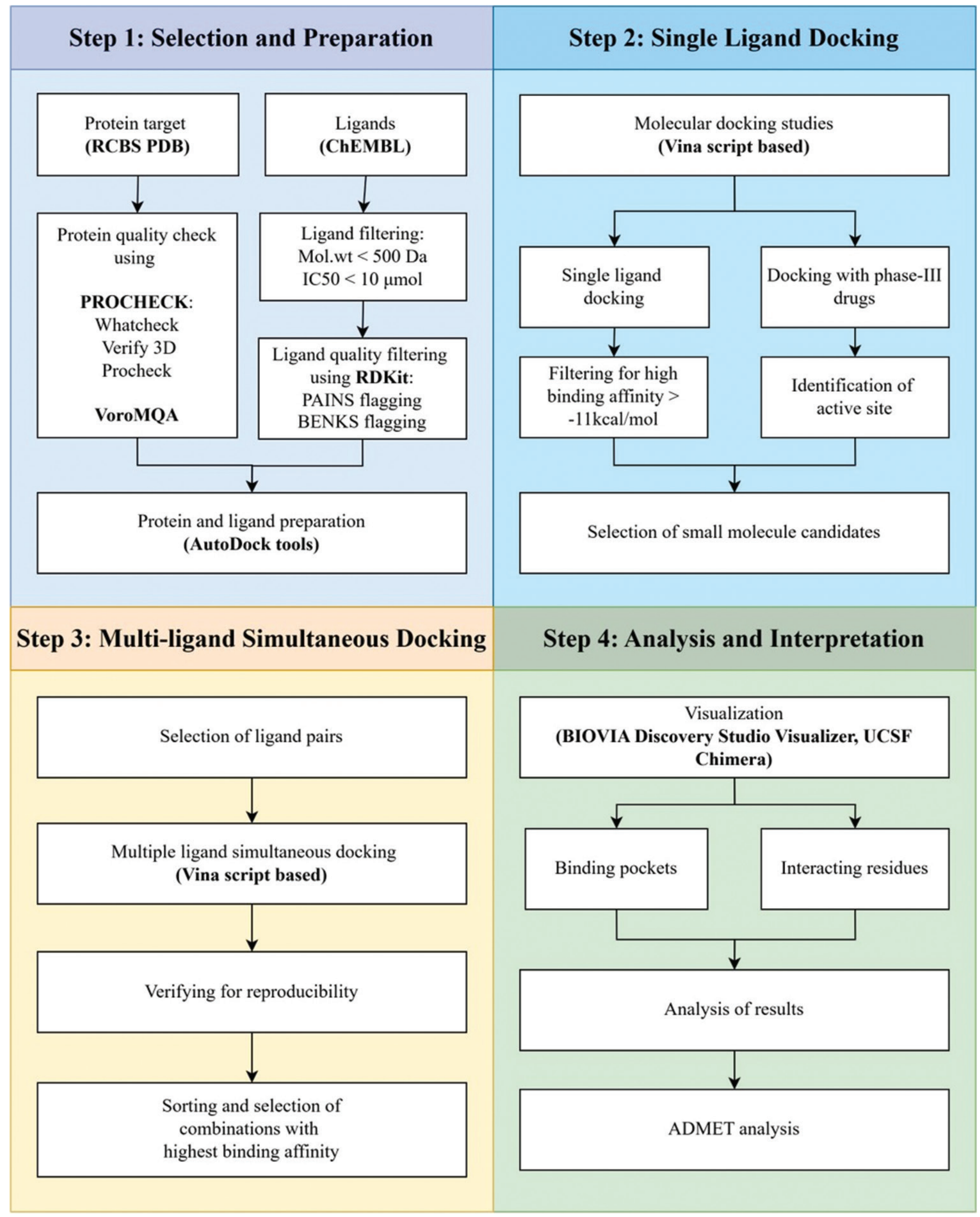

**Figure 4.** Methodological workflow illustrating the sequential steps: acquisition, filtration, and selection of candidate ligands for molecular docking, followed by docking simulations to identify potential combinations for MLSD trials. The best therapeutic combinations were subsequently subject to molecular dynamics studies.

smaller, manageable subset of high-affinity candidates for further investigation. The application of a binding affinity cutoff of −11 kcal/mol was carefully optimized to filter the most promising candidates from an initial pool of 15,641 molecules, given the computational intensity of MLSD. A higher cutoff value would have led to an unmanageable number of candidates for MLSD. For instance, using a cutoff of −10 kcal/mol would have included 110 molecules, while −9 kcal/mol would have yielded 1,292 molecules. Each additional candidate significantly increases the number of possible pairwise and multi-ligand combinations, exponentially increasing computational complexity and hindering efficient analysis. Five small molecules were identified: SM2819 (CHEMBL3695732:(5R)-2-amino-7'-(3-chlorophenyl)-4'-fluoro-3-methyl-2'-(4-methylphenyl) spiro[imidazole-5,9'-xanthene]-4-one), SM3601 (CHEMBL3695732:(5R)-2-amino-7'-(3-chlorophenyl)-4'-fluoro-3-methyl-2'-(4-methylphenyl)spiro[imidazole-5,9'-xanthene]-4-one), SM690(CHEMBL3656158:5"-methyl-6'-[3-(trifluoromethyl)phenyl]-5,6,8,9-tetrahy

dro-3'H,5''H-dispiro[-benzo[7]annulene-7,2'-indene -1',2''-[1,3,5]oxadiazole]-4''-amine), SM1953(CHEMBL4078427:6-fluoro-2-[3-[2-[6-(trifluoromethoxy)-1H-indol-3-yl]-1H-imidazol-5-yl]phenyl]-1H-benzimidazole), and SM4195(CHEMBL3586197:(1S,6R,8R,11R,12S,15S,16R,21R)-8-hydroxy-1,7,7,11,16,20,20-heptamethylpentacyclo [13.8.0.03,12.06,11.016,21] tricos-3-ene-5,19-dione).

These ligands exhibited binding affinities of −11.32 kcal/mol, −11.24 kcal/mol, −11.06 kcal/mol, −11.03 kcal/mol, and −11.02 kcal/mol, respectively. All five ligands bound to the same BACE1 pocket, but SM1953 occupied a different site and interacted with a distinct set of amino acids compared to the other four molecules (Figure 6). Several studies have investigated BACE1 inhibition. For instance, Patel *et al.*[34] reported that 28 amino acid residues in BACE1 comprise the ligand-binding site for OM99-2, including two catalytic aspartic acid residues, Asp32 and Asp228, which are essential for enzymatic activity. These residues also appear in our 2D visualization data (Table 1), confirming that our ligands interact with the same catalytic site.[34]

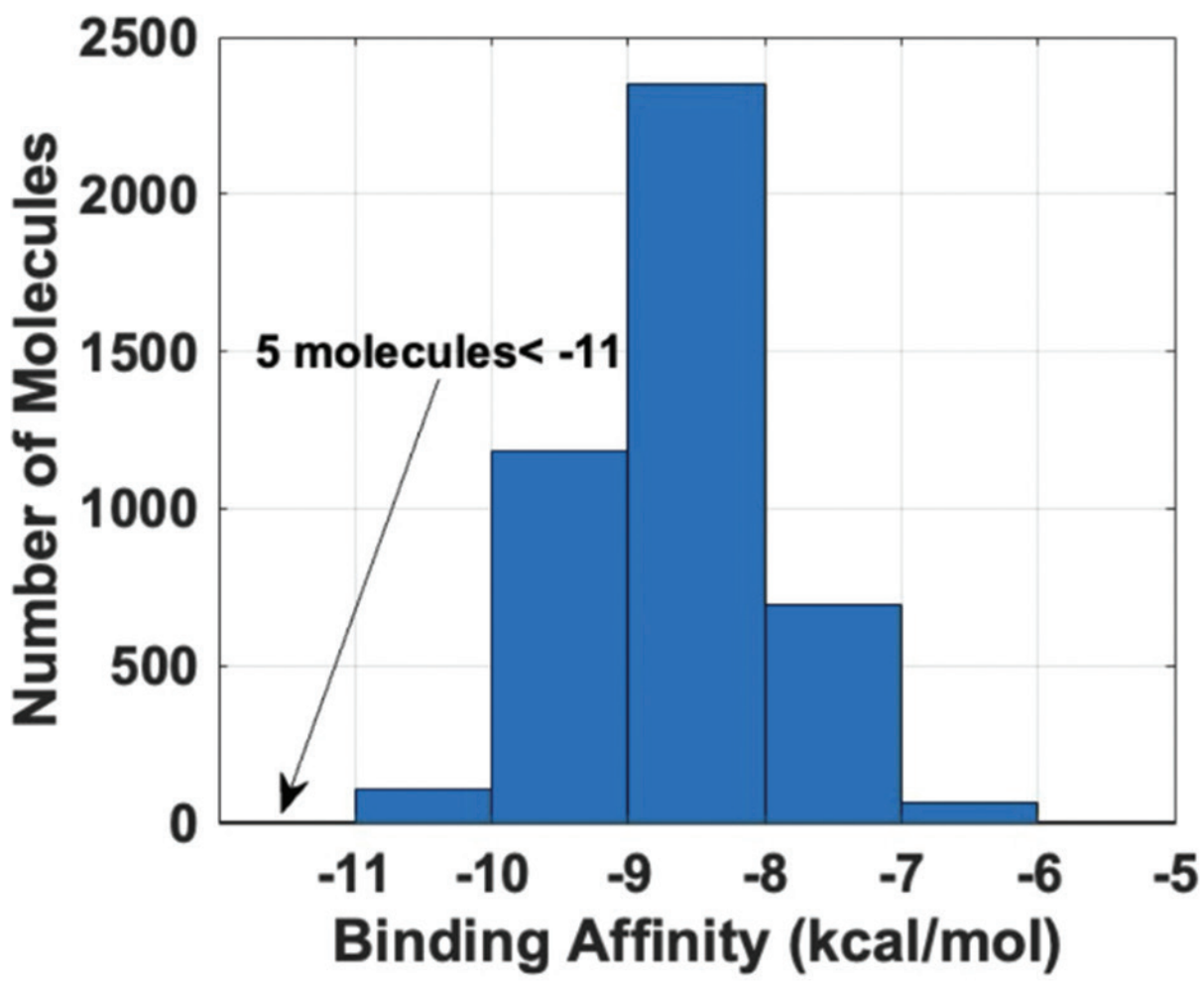

**Figure 5.** Distribution of binding affinity of molecules obtained from single-ligand molecular docking with BACE1

Multiple studies have been conducted on BACE1 inhibitors, most notably the findings from Dhanabalan et al., which identified marine bioactives such as fucotriphlorethol A and phlorofucofuroeckol that demonstrated excellent binding affinities of −17.27 kcal/mol and −13.46 kcal/mol, respectively, supported by stable MD simulations.[35] A recent docking study by Gheidari *et al.*, identified key residues in the BACE1 active site (ASP32, TYR71, LYS107, ARG128, TYR198), which are consistent with those involved in interactions in our MLSD combinations, emphasizing the robustness of our docking approach (Table 1).[36] Other studies on BACE1 include docking of naturally derived and synthetic compounds by Kalaimathi *et al.* (−11.26 kcal/mol for amygdalin) and Murad *et al.* (−8.83 kcal/mol for top MCULE ligand MCULE-3872425295-0-7), respectively.[37,38]

Furthermore, Sangeet[39] employed a machine learning-based approach to identify potential BACE1 inhibitors. Molecular docking analyses revealed detailed binding interactions, with binding energies of −8.8 kcal/mol

**Table 1. Key BACE1 binding residues identified in single docking, MLSD, and phase III inhibitors**

| Residue | Ligands from single docking | Phase III inhibitors | Ligands from MLSD |
|---|---|---|---|
| Leu30 | SM690, SM1953 | – | C1/C2/C6 (SM1953), C11 (Verubecestat) |
| Asp 32 | SM690, SM1953, SM2819, SM3601 | Atabecestat, elenbecestat, lanabecestat, verubecestat | C1/C11 (SM690), C2 (SM2819), C3 (SM3601), C6 (SM1953) |
| Tyr 71 | SM690, SM1953, SM2819, SM3601, SM4195 | Atabecestat, elenbecestat | C1/C11 (SM690), C2 (SM2819), C3 (SM3601), C6 (SM1953/Lanabecestat) |
| Gly 74 | SM4195 | – | – |
| Lys 107 | SM2819, SM3601, | - | C1 (SM690) |
| Phe108 | SM690, SM1953, SM2819, SM3601, SM4195 | Atabecestat, elenbecestat, lanabecestat, verubecestat | C1 (SM690), C2 (SM2819), C6 (SM1953/Lanabecestat) |
| Trp115 | SM4195 | Lanabecestat, verubecestat | C1/C2/C6 (SM1953), C11 (Verubecestat) |
| Arg 128 | SM690, SM2819, SM3601 | – | C1/C11 (SM690), C2 (SM2819), C3 (SM3601), C6 (Lanabecestat) |
| Tyr 198 | SM690, SM2819, SM3601, SM4195 | Atabecestat, lanabecestat, | C1/C6 (SM1953), C2 (SM2819), C3 (SM3601), C6 (Lanabecestat) |
| Asp 228 | SM690, SM4195 | Atabecestat, elenbecestat, lanabecestat, verubecestat | C1/C6 (SM1953), C11 (SM690) |

Abbreviation: MLSD: Multi-ligand simultaneous docking.

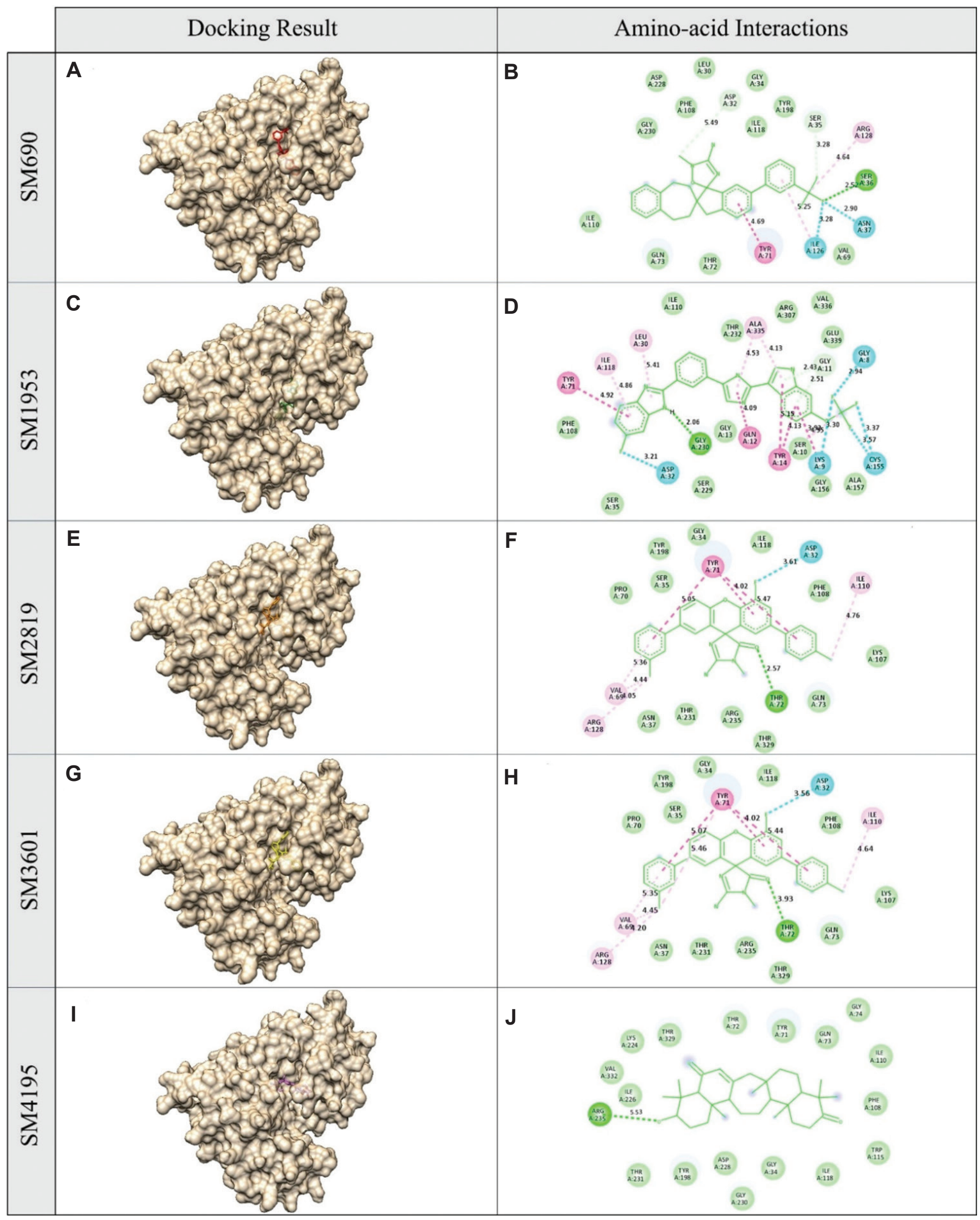

**Figure 6.** (A-J) Single ligand molecular docking results depicting the top five small molecules, their active site on β-secretase, their orientation within the binding pocket, and their respective 2D amino acid interactions within the active site

for MLC10 and −7.7 kcal/mol for MLC11. Docking was performed using AutoDock Vina following the same protocol as our study, ensuring reliable binding affinity values. Our ligands demonstrated stronger binding affinities, exceeding those reported in this study. In 2D analysis, MLC10 showed strong interactions with key BACE1 catalytic residues, including Tyr71, Gly74, Asp32, and Asp228, supporting its inhibitory potential. MLC11 engaged eight critical residues, such as Leu30, Tyr71, Phe108, and Trp115. Both compounds directly interacted with Asp32 and Asp228, suggesting potential disruption of BACE1 catalytic activity. Additional interactions with Tyr71 (flap region) and Arg128 (113s loop) indicate a multi-site binding strategy that may enhance both specificity and potency. Comparison with our 2D analysis revealed that most of our ligands target similar residues. Table 1 details the ligands interacting with key residues from each type of docking, confirming that our docking results are consistent with previous studies and that the ligands bind the correct site, reinforcing their potential inhibitory effect.[39]

The four phase III inhibitors—atabecestat, elenbecestat, lanabecestat, and verubecestat—demonstrated binding affinities of −8.01 kcal/mol, −9.05 kcal/mol, −9.43 kcal/mol, and −8.66 kcal/mol, respectively. While all phase III inhibitors bound to the same region in the pocket as the small molecules, there were slight variations in their amino acid interactions (Figure 7). Despite these differences, a common set of amino acids was involved across all ligands. Elenbecestat and verubecestat displayed similar interactions, which were somewhat distinct from those of the small molecules. On the other hand, atabecestat and lanabecestat showed amino acid interactions that were more closely aligned with the small molecules. Table 1 shows shared binding residues between our 2D inhibitor results and reported studies.

### 3.2. MLSD

In the initial trials, two-, three-, and four-ligand combinations were tested to evaluate ligand behavior in the BACE1 pocket. Two-ligand combinations proved optimal, fitting well without overlap while covering most active sites. In contrast, combinations with more than two ligands exhibited more downsides than advantages. Although the binding affinity reached −21.609 kcal/mol, this was mainly

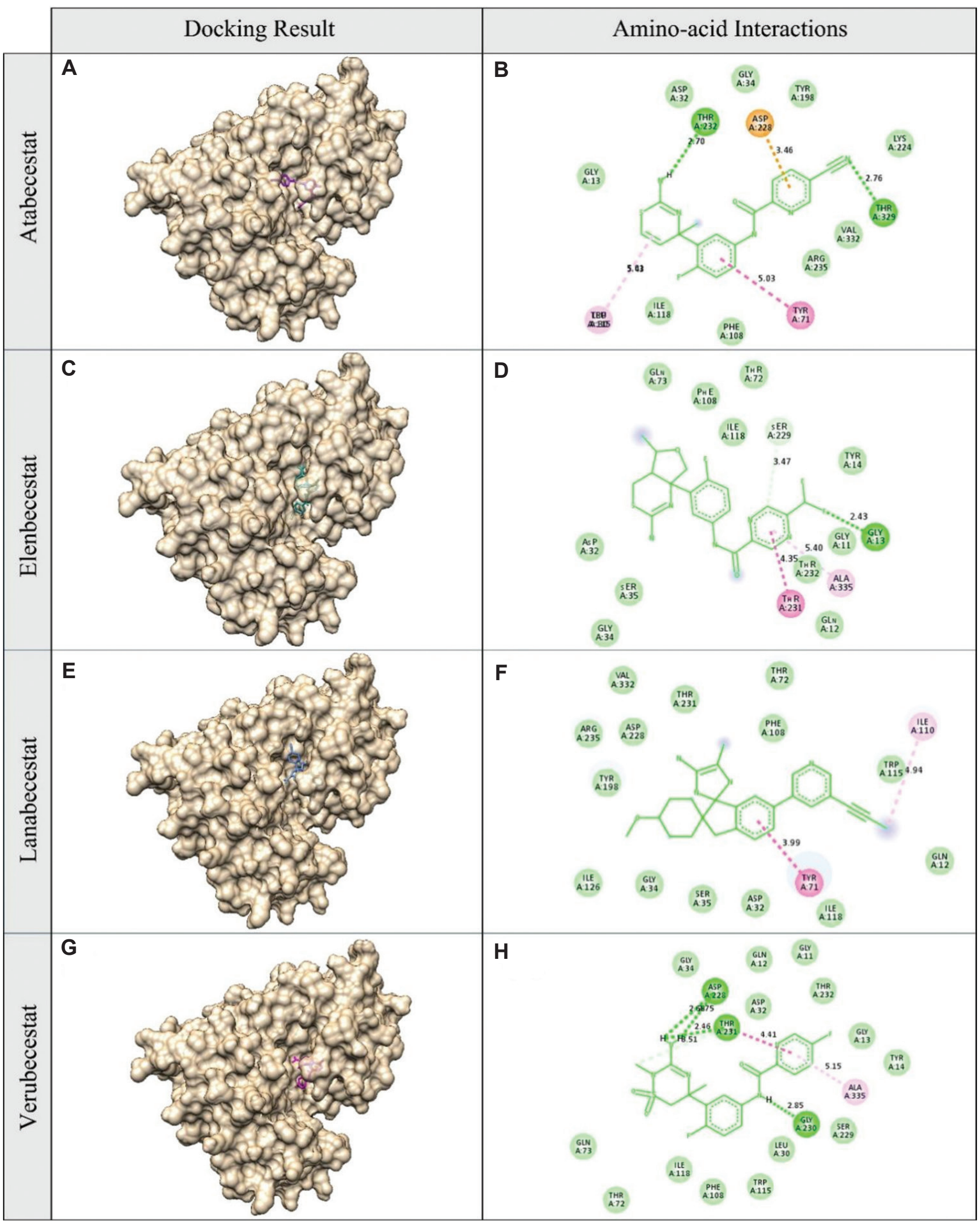

**Figure 7.** (A-H) Single ligand molecular docking results depicting the top four phase III inhibitors, their active site on β-secretase, their orientation within the binding pocket, and their respective 2D amino acid interactions within the active site

due to the addition of the third ligand rather than stronger interactions.

3D analysis revealed that SM1953 and SM2819 occupied the favorable binding site, while the third ligand, lanabecestat, bound at a secondary site (Figure A1A). Its presence disrupted the conformation of SM1953 and SM2819, leading to an unfavorable bond in SM1953 and fewer inter-ligand interactions. In 2D visualization, SM1953 formed an unfavorable bond with Cys319, likely due to steric hindrance, indicating instability (Figure A1B). Two-ligand MLSD produced six inter-ligand interactions (Figure 8B), while the three-ligand combination resulted in only three (Figure A1A), pointing to weaker cooperative binding.

As shown in Figure 9, two-ligand pairs engaged the same amino acids targeted by phase III inhibitors,

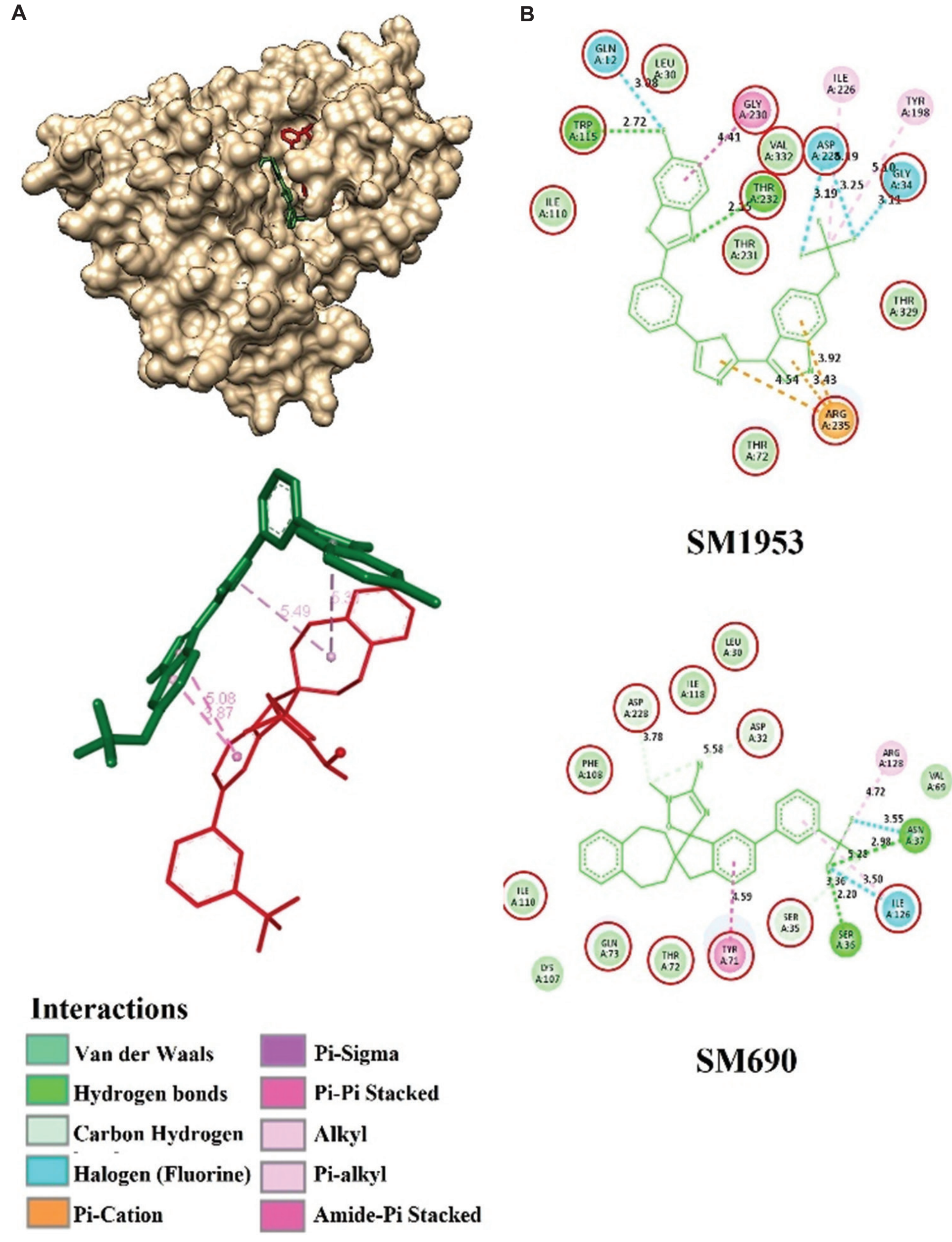

**Figure 8.** Multi-ligand simultaneous docking (MLSD) results for SM1953 (green) and SM690 (red). (A) MLSD result depicting the docked structure of SM1953 and SM690 with BACE1 protein, showing that two ligands bind to the same pocket. The inter-ligand interactions between the two small molecules are illustrated. (B) Amino acid interactions between BACE1 and the small molecules. The amino acid interactions common to the combination and phase III inhibitors are encircled in red.

making additional ligands unnecessary. A third or fourth ligand provided little extra site coverage but introduced steric overlap, competitive interactions, or displacement, reducing reproducibility and increasing computational cost. From these findings, 14 optimal ligand pairs were identified. Their interacting amino acids, spatial positions, and orientations were analyzed to ensure comprehensive coverage of the binding pocket (Table 2).

Using these combinations, MLSD was conducted according to a protocol requiring reproducibility in 3 out of 3, 4 out of 5, or 8 out of 10 trials. Combinations C4, C8, C9, C10, C12, C13, and C14 were deemed unsatisfactory due to their inability to bind to the same pocket, often resulting in one ligand typically displacing the other to a different pocket within the BACE1 protein, indicating competitive interaction. The results of these combinations have been provided in Figure A2 (Appendix). Finally, we identified 14 ligand pairs based on their interacting amino acids, positions, and orientations to ensure comprehensive binding pocket coverage through superimposed interactions. These combinations are listed in Table 2. MLSD was then performed using a protocol requiring reproducibility in 3 of 3, 4 of 5, or 8 of 10 trials. Combinations C4, C8, C9, C10, C12, C13, and C14 were unsatisfactory, as one ligand often displaced the other to a different pocket in BACE1, indicating competitive interaction. Results for these combinations are shown in Figure A2.

Furthermore, combination strategies have been explored in previous studies. For example, researchers tested the selective BACE1 inhibitor GRL-8234 in combination with the FDA-approved symptomatic drug memantine and

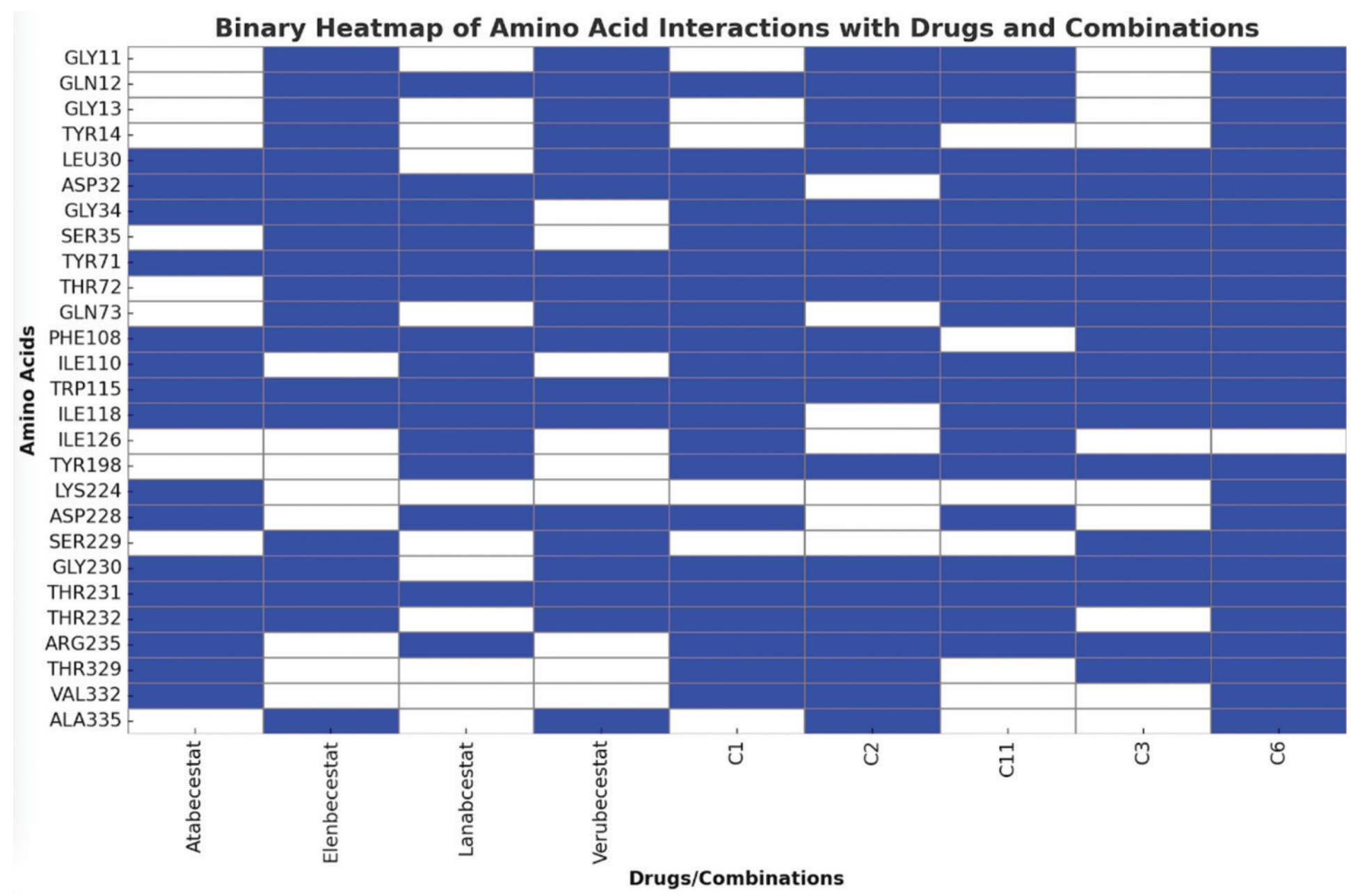

**Figure 9.** Coverage of binding pocket represented as comparison of amino acid interactions by phase III inhibitors and selected ligand combinations

**Table 2. Combinations chosen for MLSD based on their ability to comprehensively cover the binding pocket through superimposed amino acid interactions**

| Combination | Components | Combination | Components |
|---|---|---|---|
| C1 | SM1953+SM690 | C8 | Elenbecestat+SM2819 |
| C2 | SM1953+SM2819 | C9 | Elenbecestat+SM3601 |
| C3 | SM1953+SM3601 | C10 | Elenbecestat+SM4195 |
| C4 | SM1953+SM4195 | C11 | Verubecestat+SM690 |
| C5 | SM1953+Atabecestat | C12 | Verubecestat+SM2819 |
| C6 | SM1953+Lanabecestat | C13 | Verubecestat+SM3601 |
| C7 | Elenbecestat+SM690 | C14 | Verubecestat+SM4195 |

Abbreviation: MLSD: Multi-ligand simultaneous docking.

observed potential synergistic cognitive benefits within a safe dose range, highlighting the therapeutic potential of targeting BACE1 through combination strategies in AD treatment.[40] In addition, we observed that the amino acids interacting with our ligands in the MLSD combinations match the catalytic site residues reported in previous studies.[36,39] Table 1 summarizes the ligands that share these common interactions. These results indicate that MLSD, when combined with our ligand combinations, represents a promising therapeutic approach that effectively targets the BACE1 catalytic site. Finally, fragment-based approaches like those reported by Manoharan and Ghoshal[41] identified fragments with maximum binding affinities of −14.61 kcal/mol. However, MLSD combinations demonstrated superior binding affinity and interaction coverage, underscoring the potential of MLSD as a robust approach for identifying synergistic small molecule inhibitors of BACE1.[41]

Among the reproducible combinations, the top five based on cumulative binding affinities were C1 (−19.90 kcal/mol), C2 (−18.45 kcal/mol), C11 (−18.07 kcal/mol), C3 (−17.96 kcal/mol), and C6 (−17.67 kcal/mol). These combinations not only interacted with most amino acids targeted by phase III drugs but also formed additional interactions within the same pocket, contributing to higher binding affinities and broader active site coverage (Figure 9).

The stability of a ligand within a binding pocket is determined by the type and number of bonds formed between the two. These interactions include halogen (fluorine) bonds, conventional hydrogen bonds, amide-π stacking, π-anion bonds, π-π interactions, alkyl bonds, carbon-hydrogen bonds, and van der Waals forces, listed in decreasing order of strength. Halogen bonds are highly directional toward electron-rich regions and provide strong binding specificity. Hydrogen bonds orient the ligand and maintain complex stability. π interactions from aromatic rings are weaker but still contribute to the binding affinity. Van der Waals forces are the weakest and non-specific, yet their collective effect reinforces ligand-protein stability.[29]

C1 consistently bound within the same pocket across trials, despite variations in ligand orientations. The two ligands occupied adjacent positions, with SM1953 forming a hydrogen bond with THR232, similar to the phase III drug atabecestat, and SM690 forming a π-π bond with TYR71, like lanabecestat (Figure 8). These ligands target the same amino acids and form analogous bonds as established inhibitors, suggesting comparable inhibitory properties. Multiple other bonds, including halogen (fluorine) and π-π bonds, were also formed between the ligands and the protein. These non-covalent interactions significantly contribute to the overall binding affinity of the ligands. Additional bonds, including halogen (fluorine) and π-π interactions, further contributed to the overall binding affinity. Inter-ligand interactions, such as π-alkyl and π-π stacking, highlighted synergy between the molecules and likely enhanced binding by stabilizing the complexes, reducing ligand dissociation, and preventing substrate binding. Hydrophobic interactions from π-electron clouds also supported complex stability.[42]

C2 demonstrated consistent binding within the same pocket across trials, despite variations in ligand orientations. However, unlike C1, the ligands in C2 maintained their original single-ligand docking positions, occupying distinct regions within the pocket. Notably, SM2819 formed a π-π bond with TYR71, similar to lanabecestat. Ligand interaction analysis revealed multiple halogen and π bonds, which contributed to the high binding affinity. C2 also showed a higher number of inter-ligand interactions among all combinations, indicating strong synergy and enhanced complex stability. The formation of a halogen bond between the ligands further suggested the potential for a highly stable complex (Figure 10).

C11 consistently yielded the same orientation within the same binding pocket across all trials. Verubecestat formed a few hydrogen bonds, while SM690 also formed a limited number of bonds, including a π-π bond with TYR71, similar to that observed with lanabecestat. Despite the relatively few bonds, C11 exhibited a high binding affinity of −18.07 kcal/mol, primarily due to the substantial number of van der Waals interactions. Inter-ligand interactions were minimal, with only a single π-alkyl bond forming between the two ligands (Figure 11).

C3 displayed variable results, with typically one of the two small molecules forming multiple bonds with the protein, while the other interacted primarily through van der Waals forces. Nonetheless, the bonds formed were similar to those of the phase III drugs, with SM1953 binding to GLY230 with a conventional hydrogen bond like verubecestat, and SM3601 binding to TYR71 through a π-π bond like lanabecestat. Multiple inter-ligand interactions are indicative of synergism (Figure 12).

C6 included the phase III inhibitor lanabecestat in combination with SM1953. While lanabecestat formed a limited number of bonds with BACE1, SM1953 formed a conventional hydrogen bond with GLY230, similar to verubecestat, and an amide-π bond with THR231, akin to elenbecestat. In addition, SM1953 formed other strong ligand-protein interactions, including halogen bonds and π-cation bonds, contributing to the high binding affinity. C6 also demonstrated a higher number of inter-ligand interactions among all the combinations tested, highlighting the stability of the complex (Figure 13).

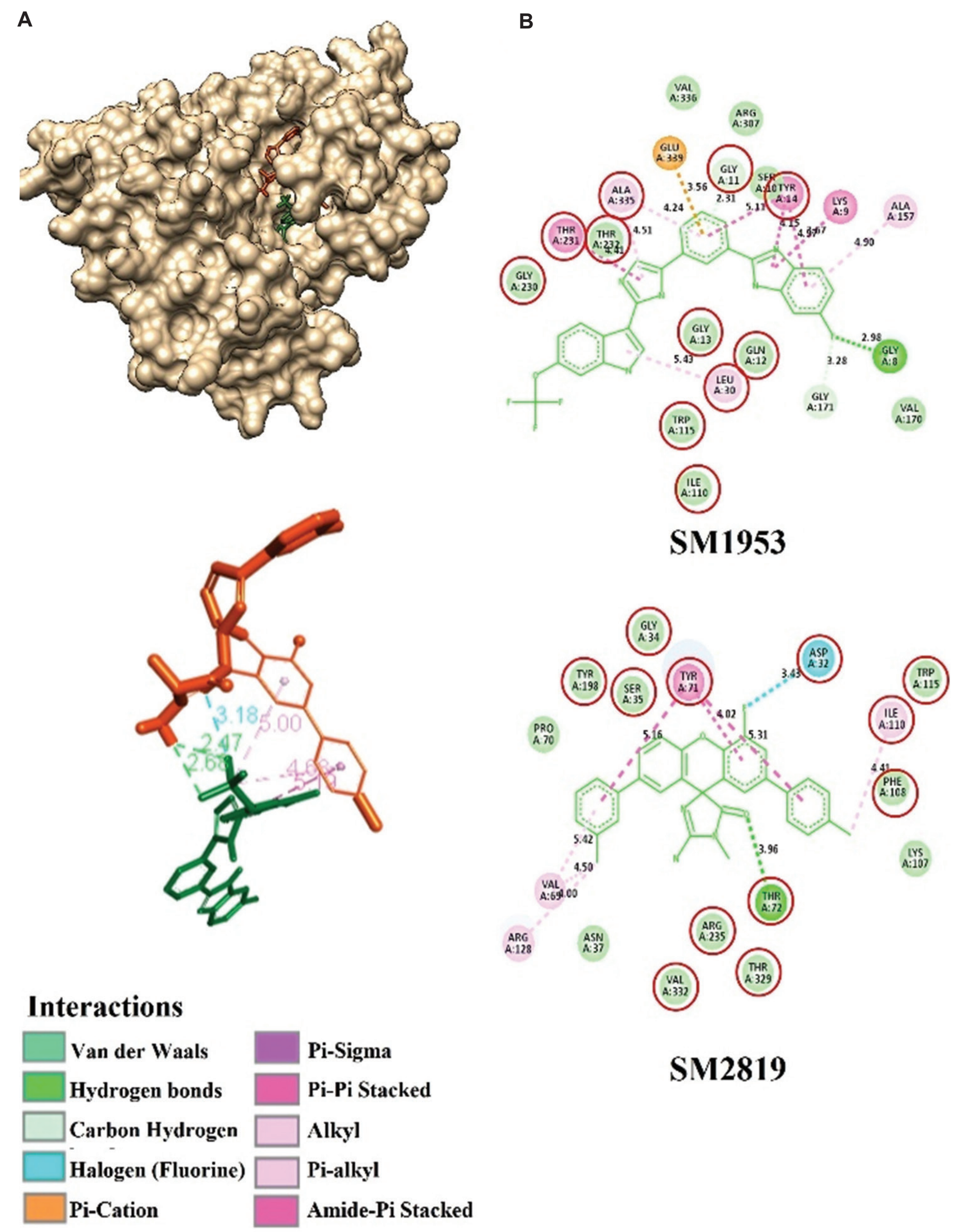

**Figure 10.** Multi-ligand simultaneous docking (MLSD) results for SM1953 (green) and SM2819 (orange). (A) MLSD result depicting the docked structure of SM1953 and SM2819 with BACE1 protein, showing that two ligands bind to the same pocket. The inter-ligand interactions between the two small molecules are illustrated. (B) Amino acid interactions between BACE1 and the small molecules. The amino acid interactions common to the combination and phase III inhibitors are encircled in red.

C1, C2, C11, C3, and C6 exhibited high binding affinities, ranging from −19.90 kcal/mol to −17.67 kcal/mol, when compared to all other combinations. However, each combination varied in ligand-ligand and ligand-protein interactions, binding affinity, consistency in reproducibility, and binding site coverage. C1 demonstrated the highest binding affinity among all combinations but had relatively fewer inter-ligand interactions and moderately lower reproducibility compared to other combinations, indicating some variability in its stability. C2, while slightly lower in binding affinity than C1, exhibited better reproducibility and inter-ligand interactions, making it a stable combination, though it did not cover the binding site as extensively as C6. C11 stood out for its excellent reproducibility and competitive binding affinity, but it lacked significant inter-ligand interactions and coverage of the binding pocket. C3 showed good inter-ligand interactions and synergistic effects, but it fell short in terms of reproducibility and binding site coverage, which limited its overall potential. C6, on the other hand, offered the most balanced performance across all key parameters. It demonstrated the highest binding site coverage, the greatest number of inter-ligand interactions, and reliable reproducibility, making it a stable combination. Although its binding affinity was slightly lower, C6 is considered a promising candidate because of its overall stability and strong binding interactions.

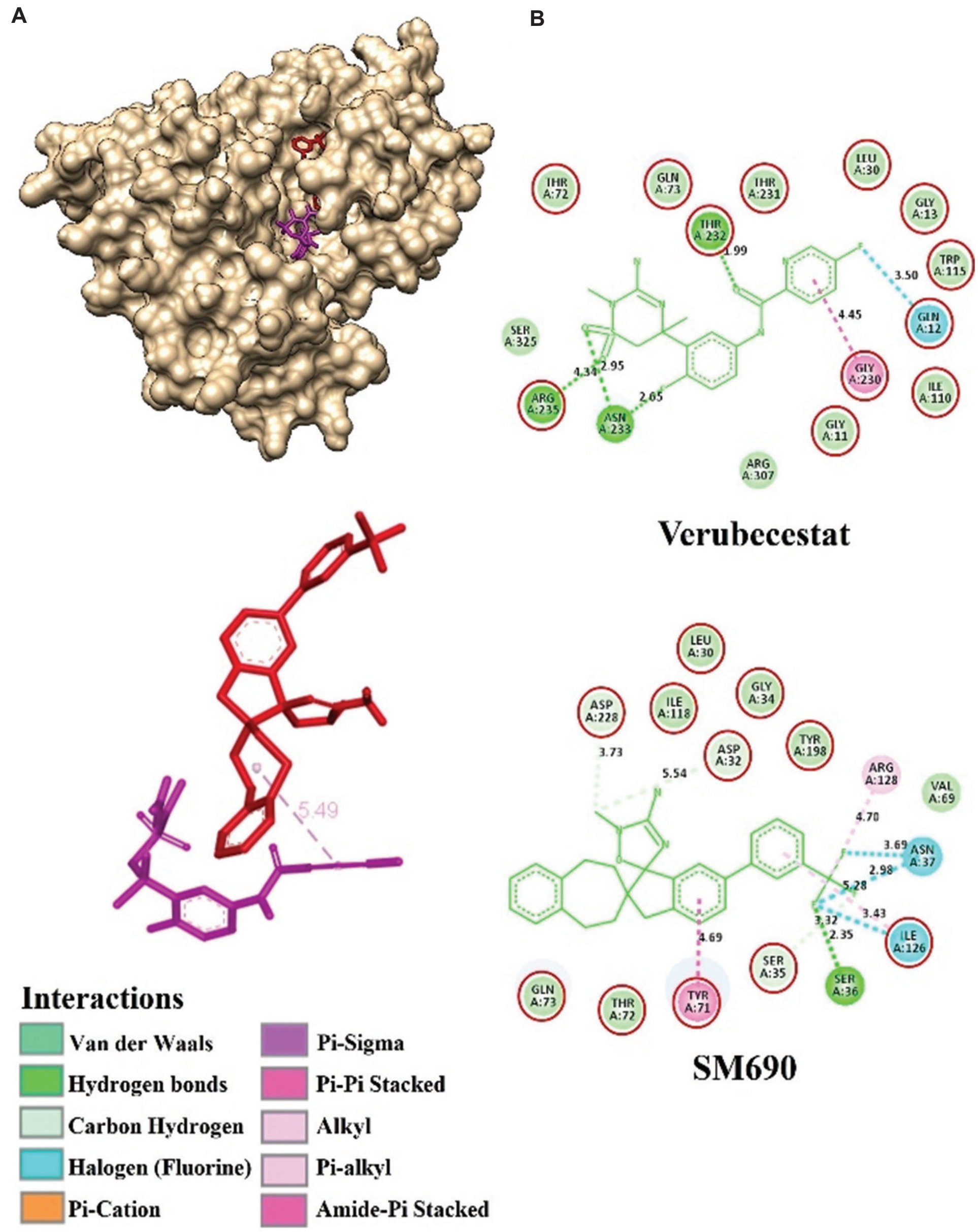

**Figure 11.** Multi-ligand simultaneous docking (MLSD) results for verubecestat (purple) and SM690 (red). (A) MLSD result depicting the docked structure of verubecestat and SM690 with BACE1 protein, showing that two ligands bind to the same pocket. The inter-ligand interactions between the two ligands are illustrated. (B) Amino acid interactions between BACE1 and the ligands. The amino acid interactions common to the combination and phase III inhibitors are encircled in red.

### 3.3. MD analysis

Following MLSD, C6 (CHEMBL4078427 and lanabecestat) was selected for MD simulation due to its broad amino acid coverage, strong inter-ligand interactions, and the synergistic pairing of a phase III drug with a small molecule adjuvant (Figure 9). It was observed that the MLSD complex exhibited lower RMSD values compared to the single docking of either SM1953 (CHEMBL4078427) or lanabecestat (Figure 14A). This reduction in RMSD highlights the superior stability of the MLSD complex, closely aligning it with the apo state. This increased stability supports the hypothesis that MLSD can enhance structural and functional properties through synergistic binding. Similarly, the Rg (Figure 14B) analysis showed that the single docking of SM1953 resulted in higher Rg values, indicating a less compact structure. In contrast, the MLSD complex demonstrated comparable Rg values to lanabecestat alone, further confirming its structural stability.

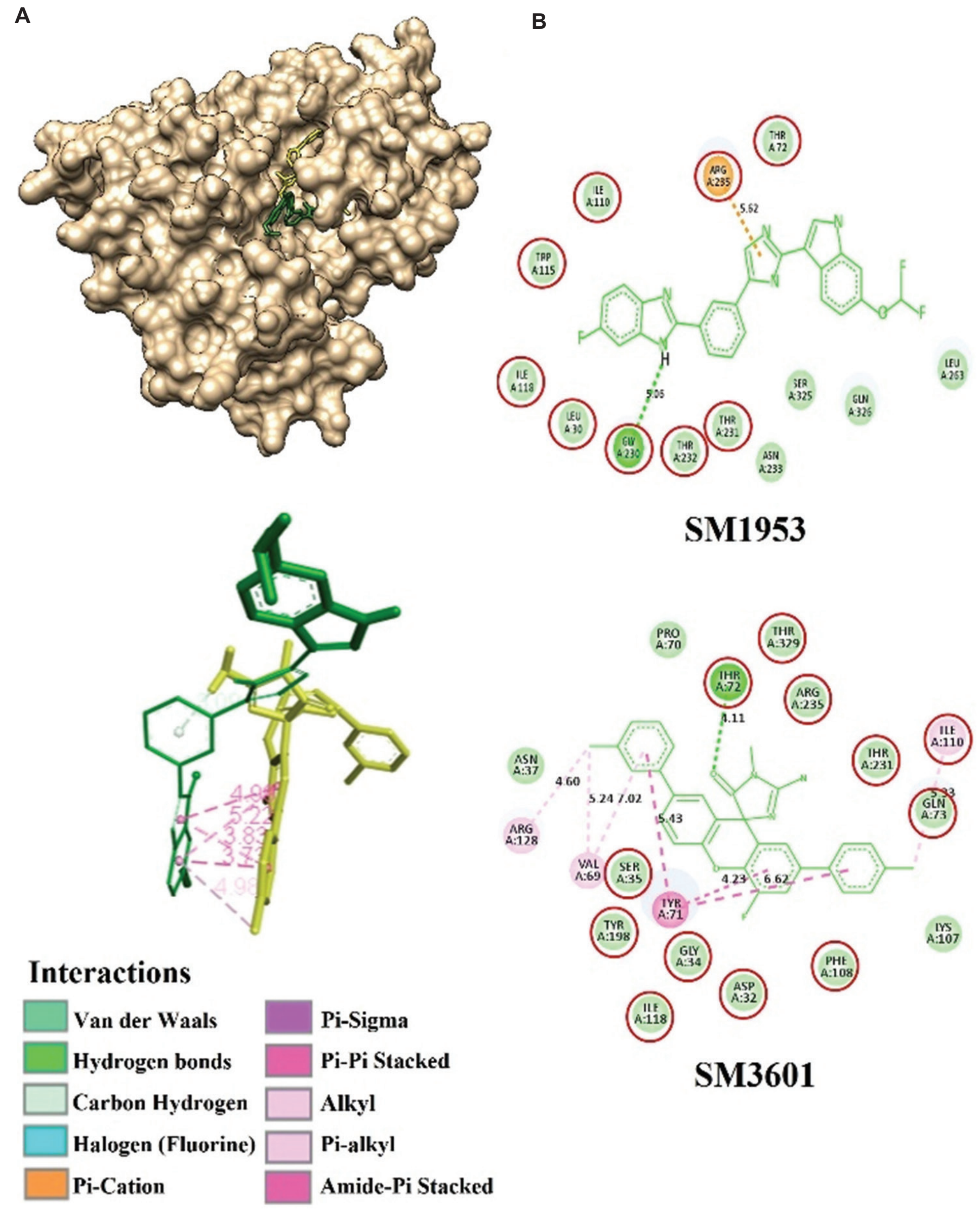

**Figure 12.** Multi-ligand simultaneous docking (MLSD) results for SM1953 (green) and SM3601 (yellow). (a) MLSD result depicting the docked structure of SM1953 and SM3601 with BACE1 protein, showing that two ligands bind to the same pocket. The inter-ligand interactions between the two small molecules are illustrated. (b) Amino acid interactions between BACE1 and the small molecules. The amino acid interactions common to the combination and phase III inhibitors are encircled in red.

Notably, the RMSD values observed in our MD simulations for the MLSD complex remained within acceptable thresholds, even as they fluctuated slightly higher than those of apo-BACE1. The apo protein showed an RMSD increase from 1.20 Å to 1.77 Å at the final frame with a maximum value of 2.32 Å while the MLSD complex increased from 1.20 Å to 2.40 Å. The Rg value of the free apo protein increased from 2.09 nm to 2.20 nm, and that of the MLSD complex increased from 2.09 nm to 2.22 nm. However, it showed a lowering of the average RMSD (1.84 Å) and Rg (2.16 nm) value of the MLSD combination compared to single ligand runs with SM1953 (RMSD = 4.29 Å, Rg = 2.38 nm) and lanabecestat (RMSD = 1.99 Å, Rg = 2.16 nm). This behavior mirrors findings from Zhang *et al.*,[43] where the RMSD for free BACE1 fluctuates between 1.53 Å and 2.21 Å, and when in complex with NB360 fluctuates between 1.44 Å and 2.25 Å, indicative of equilibrium stability despite dynamic fluctuations. Interestingly, while compounds like vasicine derivative VA10 studied by Bhanukiran *et al.*[44] exhibited promising BACE1 inhibition (−9.78 kcal/mol) with RMSD values stabilizing around 1.9 Å, our MLSD strategy achieves significantly lower binding energies.

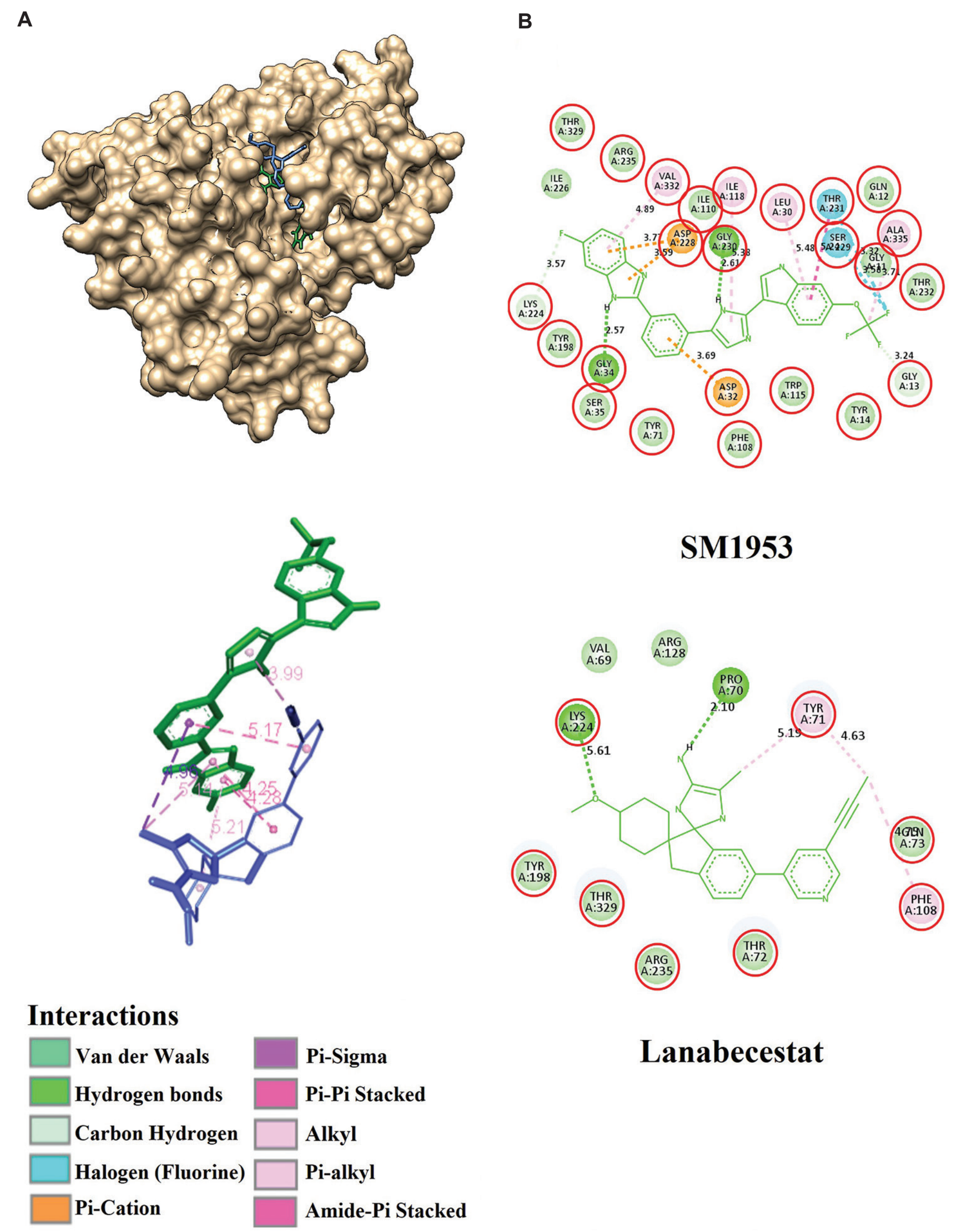

**Figure 13.** Multi-ligand simultaneous docking (MLSD) results for lanabecestat (blue) and SM1953 (green). (A) MLSD result depicting the docked structure of lanabecestat and SM1953 ligands with BACE1 protein, showing that two ligands bind to the same pocket. The inter-ligand interactions between the two ligands are illustrated. (B) Amino acid interactions between BACE1 and the ligands. The amino acid interactions common to the combination and phase III inhibitors are encircled in red.

In comparison with the study conducted by Sangeet,[39] where 150 ns MD simulations were performed with apo, atabecestat, lanabecestat, verubecestat, and MLC10, a compound generated using a machine learning model, we observed highly consistent results. In their work, the apo protein showed RMSD values between 0.15 and 0.3 nm. Similarly, in our simulations, the apo RMSD ranged from 0.15 to 0.2 nm, with the majority of values near 0.15 nm in both cases. In both studies, the apo system remained the most stable, while ligand-bound complexes displayed higher RMSD values compared to the apo. The RMSD profile of lanabecestat in our study also closely matched theirs, ranging from 0.15 to 0.3 nm in their case and 0.15–0.25 nm in ours, both centered around 0.2 nm. Likewise, for Rg, their values ranged from 2.0 to 2.2, while ours spanned 2.125–2.3. These parallels strongly support the reliability and accuracy of our MD simulations.[39] This suggests that multi-ligand docking, leveraging synergistic interactions, could potentially outperform single-molecule inhibitors in terms of binding affinity and stability.

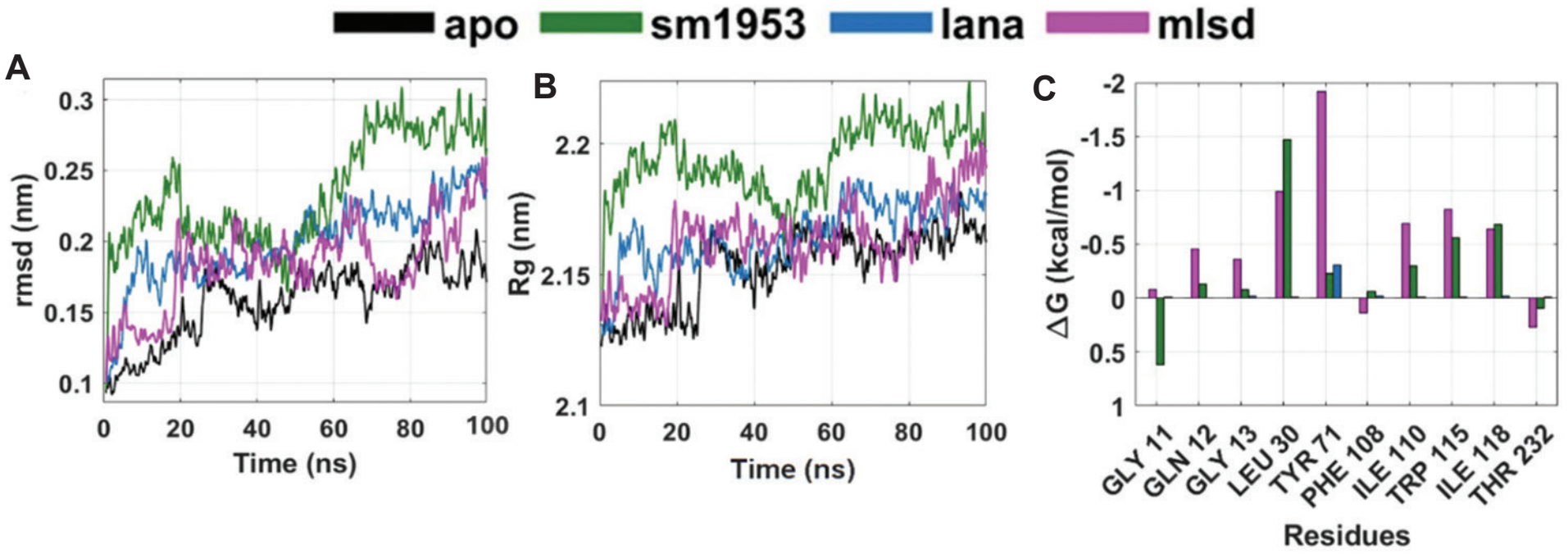

**Figure 14.** Molecular dynamics analysis of protein–ligand complexes. (A) Root mean square deviation (RMSD). (B) Radius of gyration (Rg). (C Free energy decomposition of residues for single dockings and multi-ligand simultaneous docking (MLSD) with lanabecestat and SM1953.

MM-PBSA binding free energy calculations provided quantitative validation of this enhanced stability. The MLSD complex exhibited a binding free energy of −28.11 kcal/mol, significantly outperforming the single docking energies of SM1953 (−17.07 kcal/mol) and lanabecestat (−15.22 kcal/mol). This suggested a synergistic effect, as the combined binding affinity of MLSD surpassed the additive contributions of the two individual molecules. Notably, lanabecestat, a phase III clinical inhibitor, showed the weakest binding affinity, underscoring the superior potential of the MLSD identified combination. The synergistic binding observed in MLSD is attributed to complementary interactions between SM1953 and lanabecestat. The residue-wise Gibbs free energy decomposition (Figure 14C) revealed that SM1953 formed strong interactions with residues such as LEU30 and ILE118, while the MLSD complex showed enhanced interactions with residues like GLN12, GLN13, TYR71, ILE110, and TRP115. The unfavorable interaction at GLY11, with the highest positive Gibbs free energy, indicated a potential area for further optimization.[45] This particular combination for MD simulation demonstrated the power of MLSD in identifying synergistic pairs. By combining a novel small molecule (SM1953) with a phase III inhibitor (lanabecestat), MLSD showcased its ability to identify synergistic combinations that can enhance binding and inhibitory effects. In addition, Leu30, Tyr71, Phe108, Trp115, and Thr232 were identified as key amino acid residues involved in binding (Table 1).[39]

### 3.4. ADMET analysis

ADMET analysis evaluates a drug's absorption, distribution, metabolism, excretion, and toxicity, providing insights into pharmacokinetics, safety, and efficacy. Computational *in silico* models allow early prediction of these properties, though results require experimental validation. In this study, ADMET analysis, along with molecular weight, PAINS, Brenk filters, and BBB penetration, guided the initial selection of small molecules. Molecules were shortlisted based on their ability to meet most of these criteria, ensuring a pharmacokinetically viable starting point. Table 3 lists the notable molecular and pharmacokinetic properties of the small molecule, and Table 4 lists notable toxicity characteristics of the small molecules.

Effective central nervous system (CNS) drug development faces challenges due to the protective mechanisms of the brain. ADMET properties such as the BBB, lipophilicity (LogP), hERG inhibition, and P-glycoprotein (P-gp) substrate behavior influence whether a drug can reach its target safely. Understanding how unfavorable ADMET properties impact these processes is critical for designing CNS-active compounds with optimal efficacy and safety.

The CNS is protected by the BBB, which tightly regulates the entry of drugs from the bloodstream. This barrier is formed by endothelial cells connected by tight junctions and is reinforced by metabolizing enzymes and efflux transporters, such as P-gp and multidrug-resistance proteins (MRPs), which actively remove foreign

**Table 3. Molecular and pharmacokinetic properties of the small molecules**

| Small molecule | MW (g/mol) | logP | Caco-2 (Log (cm/s)) | BBB | PPB (%) | logVDss (L/kg) | $t_{0.5}$ (h) |
|---|---|---|---|---|---|---|---|
| SM690 | 477.20 | 4.92 | −4.72 | 0.99 | 98.47 | 1.17 | 0.51 |
| SM1953 | 477.12 | 5.57 | −5.01 | 0.61 | 98.22 | 0.46 | 0.58 |
| SM2819 | 497.13 | 5.66 | −4.70 | 0.01 | 98.65 | 0.39 | 0.51 |
| SM3601 | 497.13 | 5.66 | −4.70 | 0.01 | 98.65 | 0.39 | 0.51 |
| SM4195 | 454.34 | 4.22 | −4.99 | 0.002 | 88.45 | 0.16 | 0.52 |

Abbreviations: BBB: Blood–brain barrier; MW: Molecular weight; PPB: Plasma protein binding.

**Table 4. Toxicity profile of the small molecules**

| Small molecule | DILI | SkinSen | Carcinogenicity | Respiratory | Neurotoxicity-DI | Nephrotoxicity-DI | Genotoxicity |
|---|---|---|---|---|---|---|---|
| SM690 | 0.3209 | 0.2789 | 0.3046 | 0.7344 | 0.9402 | 0.7407 | 1.0000 |
| SM1953 | 0.9950 | 0.0026 | 0.7749 | 0.9818 | 0.9880 | 1.0000 | 1.0000 |
| SM2819 | 0.9902 | 0.4793 | 0.4444 | 0.4049 | 0.9960 | 0.9835 | 0.9996 |
| SM3601 | 0.9902 | 0.4793 | 0.4444 | 0.4049 | 0.9960 | 0.9835 | 0.9996 |
| SM4195 | 0.3640 | 0.8991 | 0.8784 | 0.7767 | 0.2437 | 0.8827 | 0.8716 |

Abbreviations: DI: Drug interaction; DILI: Drug-induced liver injury.

compounds. Consequently, many drugs that are effective *in vitro* or in animal studies fail to reach their CNS targets in humans, contributing to the high attrition rate in CNS drug development. To address this, early assessment of BBB permeability using efficient, high-throughput approaches is essential, reducing the dependence on costly animal experiments and minimizing the risk of failure in clinical trials.[46]

Resistance to CNS drugs can also arise when the BBB prevents sufficient drug entry. P-gp, an efflux transporter in the BBB, pumps out foreign substances, including therapeutic drugs, which can limit their effectiveness. While P-gp inhibition may enhance brain drug delivery, it may also increase the risk of toxicity by allowing harmful substances to accumulate. Therefore, determining whether a drug is a P-gp substrate is an important step in early CNS drug development.[47]

High lipophilicity (LogP) often results in compounds with low water solubility, poor absorption, and rapid metabolic breakdown, which can reduce their effectiveness. Highly lipophilic molecules are also more likely to bind to unintended protein targets, increasing the risk of toxicity. In the context of CNS drug development, lipophilicity is particularly important because drugs must cross the lipid-rich BBB. A compound that is too lipophilic may accumulate in fatty tissues or be metabolized too quickly, while one that is too hydrophilic may fail to penetrate the brain. Therefore, achieving balanced lipophilicity is essential for effective CNS drug delivery.[48]

The hERG channel is a potassium-selective ion channel critical for cardiac repolarization. If a drug blocks the hERG channel, the QT interval on an electrocardiogram (ECG) can become prolonged, leading to long QT syndrome (LQTS) and dangerous heart rhythm disturbances such as ventricular fibrillation or tachycardia. Because many drugs, including CNS-targeted compounds, can unintentionally interact with hERG, early prediction using computational models is crucial to avoid severe cardiac side effects.[49] Overall, unfavorable ADMET properties can significantly influence the safety and efficacy of CNS drugs. Early identification and careful optimization of these properties are crucial for improving the likelihood of a compound to reach its CNS target to exhibit the therapeutic effects and maintain human safety.

SM690, with a logP of 4.19, demonstrated a balance between lipophilicity and aqueous solubility, making it suitable for membrane permeability. SM690 exhibited high Caco-2 permeability (−4.72 log units), which predicted good intestinal absorption and suitability for oral administration. However, as both a P-gp substrate and inhibitor, SM690 both undergoes and modulates efflux. While its dual role could enhance distribution across membranes like the BBB, it also posed a risk of efflux-related variability and drug–drug interactions (DDIs). High plasma protein binding (98.5%) decreased free drug availability but conferred a longer half-life and sustained action. SM690 showed significant distribution, with a high volume of distribution (Vd) and strong BBB penetration, making it a promising candidate for CNS targets. Furthermore, its inhibition of MRP1 highlighted potential interactions with chemotherapeutic agents such as etoposide and antibiotics like levofloxacin and certain endogenous compounds like bilirubin, possibly enhancing their efficacy or toxicity. Its metabolism through CYP1A2, CYP2C19, and CYP3A4 results in moderate clearance (7.57 mL/min/kg). However, the prediction of hERG blockade indicates a risk of QT prolongation, warranting caution in cardiac patients or with concomitant QT-prolonging drugs such as class Ia antiarrhythmics, macrolides, and certain antidepressants such as citalopram and amitriptyline. SM1953 demonstrated a higher logP of 5.5, correlating with poor aqueous solubility (logS −6.085) and high lipophilicity, as indicated by its high pKa of 13.034. Regardless, advanced formulation strategies, such as lipid-based or nanoformulations, may be required to enhance its oral bioavailability. Its moderate permeability is reflected in Caco-2 and MDCK scores of −5.013 and −4.821 log units, respectively. Although SM1953 has qualified multiple drug-likeness rules, its oral absorption remained limited due to low solubility. SM1953 is also a potent P-gp inhibitor, increasing the likelihood of DDIs, particularly with P-gp substrates like digoxin. High plasma protein binding and moderate log VDss (0.46) suggested limited free drug availability and tissue distribution, with lower BBB penetration compared to SM690. Its inhibition of OATP1B1/1B3 could result in reduced hepatic clearance of substrates such as statins, methotrexate, and bilirubin, increasing systemic exposure and toxicity.

Metabolism through CYP1A2 and CYP2D6, alongside inhibition of CYP2C19 and CYP2C9, suggested a potential for enzyme-mediated DDIs. Despite lower plasma clearance (4.8 mL/min/kg) than SM690, possibly due to its low renal clearance owing to its mostly unionized state at physiological pH, SM1953 has a similar half-life. Like SM690, its predicted hERG blocking property raises concerns for QT prolongation, emphasizing the need for careful monitoring. SM2819 and SM3601 shared a high logP of 5.65, which resulted in greater lipophilic propensity. It is predicted to exhibit moderate permeability with Caco-2 scores of −4.7 log units. Despite adequate membrane permeability, their likelihood of being P-gp substrates raised concerns about efflux at the intestinal epithelium and BBB, reducing bioavailability. Both candidates are inhibitors of OATP1B1 and MRP1, which may have increased tissue concentrations of co-administered drugs, prolonging exposure and heightening the risk of toxicity, especially for drugs with narrow therapeutic windows. SM4195 demonstrated favorable ADMET properties, with a logP of 4.22 supporting good permeability and moderate solubility. Its Caco-2 permeability of −4.99 log units indicated moderate intestinal absorption and oral uptake. Unlike the other candidates, SM4195 does not interact with P-gp, reducing the risk of efflux-related DDIs and variability. Its lower plasma protein binding increased free drug availability (15.8%), though its limited BBB permeability reduced its potential for CNS applications. Clearance is primarily hepatic, as SM4195 is metabolized by CYP1A2, CYP2C19, and CYP3A4, with a higher plasma clearance (11.88 mL/min/kg) compared to other candidates. Notably, SM4195 exhibited a lower toxicity profile, making it a promising candidate for further development despite its limited CNS applicability.

Regarding toxicity, each compound presented certain system-specific toxicities, notably SM1953, which displayed elevated values in drug-induced liver injury (DILI), neurotoxicity, nephrotoxicity, and genotoxicity, indicating significant safety concerns. A value of 1.0 denoted the major concern for these toxicities. However, it is important to note that if these small molecules are used in combinational therapies, the required dosage to achieve therapeutic efficacy could be significantly reduced due to the synergistic effects of the ligands, effectively reducing the adverse effects of these drug candidates.

Taken together, while the ADMET analysis predicts the ligands to be BCS Class 2 drugs, each candidate presents distinct advantages and limitations. SM690 demonstrated balanced permeability and distribution but posed risks due to transporter interactions and hERG blockade. SM1953 required advanced formulation strategies and has significant transporter-mediated DDIs. SM2819 and SM3601 face challenges with suboptimal pharmacokinetics and toxicity profiles, necessitating significant structural modifications to improve their usability. SM4195 stands out for its favorable toxicity profile and oral absorption but requires modifications for CNS targeting. Further, *in vivo* studies are necessary to refine these findings and optimize therapeutic efficacy.

## 4. Limitations and future directions

The present study provides significant insights into the potential of small molecule-based MLSD as a strategy to enhance BACE1 inhibition. We comprehensively screened 15,641 small molecules, identifying five small molecules with exceptional binding affinities. Combinations of these small molecules and four phase III inhibitors were docked simultaneously. Out of 14 combinations, six demonstrated superior binding affinities and dynamic stability compared to single-ligand inhibitors. Among the combinations tested, SM1953 and lanabecestat emerged as particularly promising, with a binding affinity of −17.67 kcal/mol. This combination not only exhibited superior inter-ligand synergy but also demonstrated improved structural stability and compactness during MD simulations, as indicated by lower RMSD and Rg values compared to single-ligand inhibition. Although these results offer a preliminary framework for drug discovery and show promising potential, further *in vitro* and *in vivo* validation is essential. Many previous candidates in this field have been discontinued due to harmful side effects, highlighting the need for cautious evaluation.

For instance, a study has identified isophthalic acid derivatives containing imidazole and indole groups as potent BACE1 inhibitors. Using a structure-activity relationship (SAR) approach, the researchers developed compounds exhibiting strong enzyme and cellular inhibition. One of the compounds, SM1953, showed BACE1 inhibition with an $IC_{50}$ of 120 ± 9 nM, indicating strong but slightly lower potency than some analogs. The value was determined through a biochemical enzyme inhibition assay. In a cell-based ELISA assay, SM1953 achieved an $EC_{50}$ of 0.29 μM with 87% inhibition, demonstrating the best cellular efficacy. Its polar surface area (82.4 Å$^2$) and high ClogP (6.42) suggest limited BBB penetration and potential solubility or toxicity issues. Although SM1953 combines strong enzyme and cellular activity, its physicochemical properties require optimization for effective CNS drug development.[50]

Similarly, another study by Nguyen *et al.*[51] reported thirteen compounds isolated from *Lycopodiella cernua*, a Chinese herbal medicine. Structural analysis using nuclear magnetic resonance, mass spectrometry, and Mosher's method confirmed that these triterpenoid derivatives possessed either an oleanane-type or serratene-type core skeleton, along with one new hydroxy unsaturated fatty acid similar to SM4195. These triterpenoid derivatives from *L. cernua* exhibited strong inhibitory activity against key neural enzymes, including β-secretase (BACE1), acetylcholinesterase (AChE), and butyrylcholinesterase (BChE), with $IC_{50}$ values ranging from 0.22 to 0.42 μM. These submicromolar values indicated potent inhibition and suggest potential in preventing APP cleavage and modulating cholinergic activity. Overall, the compounds demonstrated strong multitarget activity, highlighting their promise as neuroprotective agents for AD therapy.[51]

Although no data are available for SM2819 and SM3601, a patent described a class of compounds with a common backbone designed to inhibit BACE (β-site APP cleaving enzyme), which generated Aβ peptides implicated in AD and other CNS disorders. Structural variations at specific positions give rise to related sub-compounds. The patent covered the compounds, their formulations, therapeutic applications, and synthesis methods. Modifications with functional groups can enhance selective activity, while pharmacokinetic and pharmacodynamic properties determine overall effectiveness. The activity of these compounds can be demonstrated in both *in vitro* and *in vivo* studies.

Compound 33, (4R)-2-amino-7'-(3-chlorophenyl)-4'-fluoro-1-methyl-2'-(4-methylphenyl)spiro[imidazole-4,9'-xanthen]-5(1H)-one, shared the same spiroimidazole-xanthene backbone. It exhibited strong BACE1 inhibition with an FRET IC50 value of 0.0041 μM and HEK cell $IC_{50}$ value of 0.237 μM, indicating efficient active-site binding and cellular activity, although some potency is reduced by limited permeability or metabolism. This backbone tolerates aromatic substitutions while maintaining high biochemical potency and measurable cell activity, making it a promising lead. High molecular weight, lipophilicity, and stereochemical sensitivity may limit brain penetration, but similar compounds are expected to retain activity if key substituents and stereochemistry are preserved.[52] Similarly, SM690 and related compounds in the series show $IC_{50}$ values below 500 nM, with many below 100 nM, reflecting very high potency. Only a few exhibit micromolar activity (1–10 μM), indicating moderate-to-weak inhibition. Overall, SM690 and its analogs demonstrated strong submicromolar inhibition, consistent activity, and a well-optimized chemical framework, providing a promising basis for further lead development.[53]

Apart from small molecules, several phase III inhibitors also showed certain disadvantages, which led to the discontinuation of their clinical trials. For instance, lanabecestat is an acyl guanidine BACE1 inhibitor. Phase I trials showed good safety and metabolism, and phase II/III results indicated improved cognition and reduced Aβ42 levels in cerebrospinal fluid (CSF). However, trials were discontinued due to weight loss, skin depigmentation, and psychiatric effects. Verubecestat exhibited an $IC_{50}$ value of 2.2 nM and showed strong activity through amidine interaction with the BACE1 catalytic diad and reduced P-gp efflux due to its *iminothiadiazinane* core. Despite favorable pharmacokinetics, it failed to improve cognition and was stopped in phase III because of adverse effects such as rashes. In phase I, elenbecestat resulted in 92% reduction of Aβ42 levels in CSF, consistent with phase II results. However, it was later discontinued due to side effects such as dizziness, nightmares, elevated liver enzymes, and hippocampal atrophy. Atabecestat demonstrated dose-dependent CSF Aβ reduction by 50–90% in phase II but was halted due to liver enzyme abnormalities.[34]

Despite advances in drug discovery, repeated failures necessitate new approaches to treating AD. Combination therapy is one such strategy, targeting multiple sites simultaneously to enhance inhibitory effectiveness. This allows for faster drug action with a shorter duration, potentially reducing effects on non-target proteins and organs, thereby minimizing side effects. The patent for SM1953 highlights that the compounds of the invention can be administered alone or in combination with other therapeutic agents. In combination therapy, agents may be given simultaneously or sequentially, either as separate formulations or as a single composition with fixed ratios. These combinations may include therapies targeting beta-secretase, gamma-secretase, or other factors influencing amyloid-beta formation and plaque deposition. The invention also allows flexible dosing and administration sequences. Variations and modifications apparent to skilled individuals are considered within the scope of the invention.[52]

Thus, the identified ligand combinations may offer a more practical approach by balancing efficacy with reduced risk. Supported by MLSD and MD simulations predicting stability, binding affinities, and conformational behavior, this study provides a computational framework for BACE1 inhibitor design. Assessing the stability of ligand-protein complexes in physiological environments will further clarify their clinical relevance and help bridge the gap from theoretical predictions to therapeutic applications. Nonetheless, *in vitro* and *in vivo* validation is essential to confirm biological activity, pharmacokinetics, and safety, as factors such as uptake, bioavailability, and metabolic stability will ultimately determine therapeutic potential. Furthermore, refining the ADMET profiles of these ligand combinations is essential to evaluate their pharmacokinetic properties and safety, although experimental validation is crucial to fully understand their behavior within complex biological environments.

## 5. Conclusion

The complicated pathophysiology, stringent pharmacokinetic requirements, trial design difficulties, and high failure rates of AD make the discovery of new medications for these neurodegenerative diseases extremely difficult. These difficulties highlight how crucial it is to use the available data to create novel strategies, such as combination medicines, that may provide more potent therapy options. This study aimed to identify ligand combinations that work well together to inhibit BACE1 through MLSD. This strategy offers a viable means of creating an effective combination treatment for AD. The integration of small molecules with phase III inhibitors like CHEMBL3656158 with verubecestat and CHEMBL4078427 with lanabecestat highlights the potential of small molecules to enhance the inhibitory efficacy of these drugs in combination therapies. This study presents new avenues for developing pharmacotherapies for diseases with limited treatment options. Furthermore, the results underscore the potential of multi-drug therapies and fragment-based drug design strategies to overcome drug resistance, optimize dosages, and mitigate side effects. Future directions for this research include conducting MD simulations and validating findings through wet lab studies using cell line assays. This integrated approach will strengthen the reliability and applicability of the results. Advancements in this field could potentially lead to life-saving multidrug therapies that significantly improve the quality of life for individuals affected by AD.

## Acknowledgments

The authors thank AIC-DSU for supporting them with the server for conducting the study.

## Funding

None.

## Conflict of interest

The authors declare that they have no competing interests.

## Author contributions

*Conceptualization*: Pronama Biswas, Surya Shanbhog, Belaguppa Manjunath Ashwin Desai
*Formal analysis*: All authors
*Funding acquisition:* Pronama Biswas, Belaguppa Manjunath Ashwin Desai
*Investigation*: All authors
*Methodology*: Surya Shanbhog, Merla Sudha
*Writing–original draft:* Surya Shanbhog
*Writing–review & editing:* All authors

## Ethics approval and consent to participate

Not applicable.

## Consent for publication

Not applicable.

## Availability of data

The data generated and analyzed during the current study are available from the corresponding author on reasonable request.

## Appendices

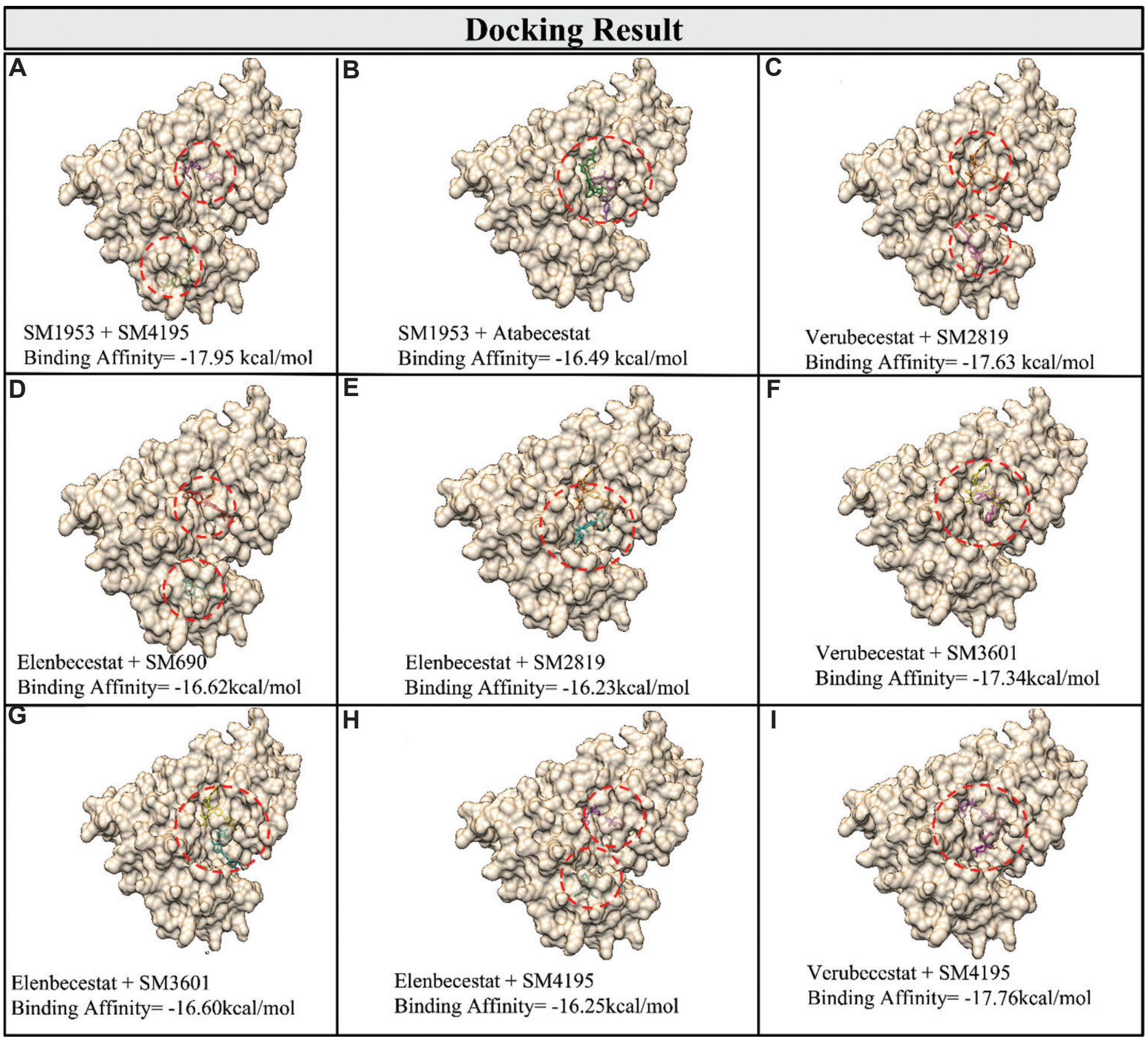

**Figure A1.** (A-I) Overview of the nine ligand combinations tested through multi-ligand simultaneous docking (MLSD) that did not meet the reproducibility criteria for successful synergistic binding. Each combination was evaluated based on the consistency of binding affinity improvements across multiple docking trials, with reproducibility defined as consistent results in at least 3 out of 3, 4 out of 5, or 8 out of 10 trials. These combinations, while not selected for further molecular dynamics simulations, provide valuable insights into the interaction landscape and binding variability among screened ligand pairs.

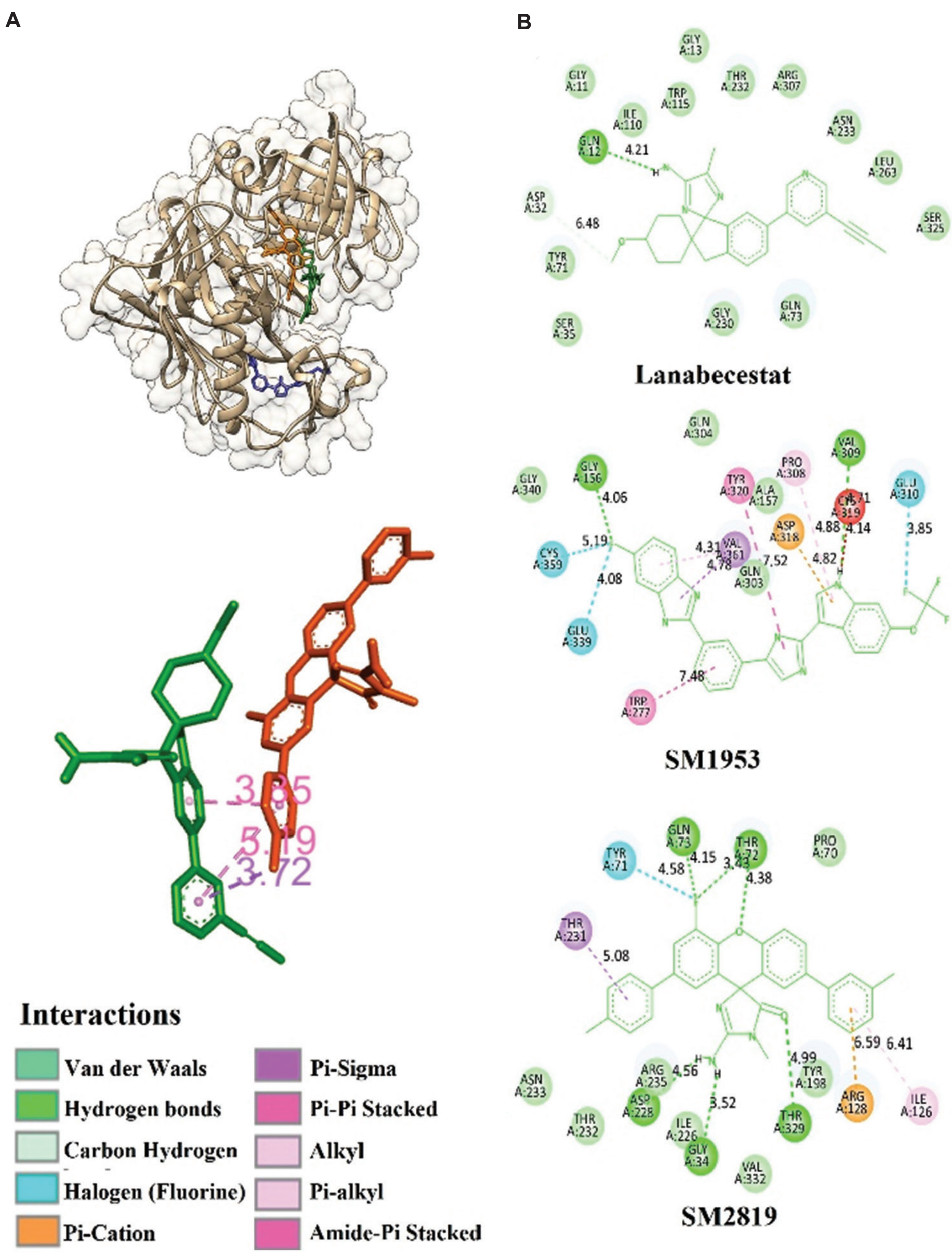

**Figure A2.** Multi-ligand simultaneous docking (MLSD) results for lanabecestat (blue), SM1953 (green), and SM2819 (red). (A) Docked structures of lanabecestat, SM1953, and SM2819 with the BACE1 protein. SM1953 and SM2819 bind to the same pocket, while lanabecestat occupies a second pocket. Inter-ligand interactions between SM1953 and SM2819 are highlighted. (B) Amino acid interactions between BACE1 and the small molecules.